\documentclass[a4paper,english,conference]{ieeeconf}

\usepackage[utf8]{inputenc}
\usepackage[hidelinks]{hyperref}
\usepackage{amsmath,amsfonts}
\usepackage[font=footnotesize]{subcaption}
\usepackage[font=small,labelfont=bf]{caption}
\usepackage{cleveref}
\usepackage{pgfplots}
\usepackage{pgfplotstable}
\usepackage{algpseudocode}
\usepackage{bm}
\usepackage{algorithm}
\usepackage{csvsimple}
\usepackage{bbm}
\usepackage{siunitx}
\usepackage{nicefrac}
\usepackage[prependcaption=inline]{todonotes}
\usepackage{microtype}
\usepackage{lipsum}

\RequirePackage{pgfplots,pgfplotstable}
\pgfplotsset{compat=newest}
\usepgfplotslibrary{units,groupplots,colormaps,fillbetween}

\usepackage[absolute]{textpos}
\newcommand{\copyrightstatement}{
	\begin{textblock*}{17cm}(20mm,1mm)    
		\noindent
		\footnotesize
		\copyright 2019 IEEE. Personal use of this material is permitted. Permission from IEEE must be
		obtained for all other uses, in any current or future media, including
		reprinting/republishing this material for advertising or promotional purposes, creating new
		collective works, for resale or redistribution to servers or lists, or reuse of any copyrighted
		component of this work in other works. DOI: 10.1109/ITSC.2019.8917384
	\end{textblock*}
}

\usepackage{tikz}

\usetikzlibrary{
	arrows,
	arrows.meta,
	angles,
	babel,
	backgrounds,
	calc,
	decorations,
	decorations.pathreplacing,
	decorations.pathmorphing,
	fit,
	intersections,
	matrix,
	petri,
	positioning,
	shapes,
	spy,
	through,
	quotes,
}

\newcommand{\eg}{e.\,g.,\ }

\DeclareMathOperator*{\argmin}{arg\,min}

\IEEEoverridecommandlockouts

\begin{document}
\copyrightstatement
\bstctlcite{bibcontrol_etal4}

\title{\LARGE \bf%
    Fast 3D Extended Target Tracking
    \\ using NURBS Surfaces
}

\author{
    Benjamin Naujoks%
    \thanks{
        All authors are with the Institute for Autonomous Systems Technology (TAS) of the Universität der Bundeswehr Munich, Neubiberg, Germany. Contact author email: benjamin.naujoks@unibw.de},
    Patrick Burger
    and %
    Hans-Joachim Wuensche
}

\maketitle
\begin{abstract}
    This paper proposes fast and novel methods to jointly estimate the target's unknown 3D shape and dynamics.
Measurements are noisy and sparsely distributed 3D points from a light detection and ranging (LiDAR) sensor.
The methods utilize non-uniform rational B-splines (NURBS) surfaces to approximate the target's shape.
One method estimates Cartesian scaling parameters of a NURBS surface, whereas the second method estimates the corresponding NURBS weights, too.
Major advantages are the capability of estimating a fully 3D shape as well as the fast processing time.
Real-world evaluations with a static and dynamic vehicle show promising results compared to state-of-the-art 3D extended target tracking algorithms.

\end{abstract}
\section{Introduction}
Target tracking is an essential requirement for enabling autonomous driving.
For instance, the autonomous vehicle has to react to other dynamic objects, \eg cars or pedestrians.
Typically, the estimated kinematic values are Cartesian position and velocity, and the most common approach is the small-target-assumption or point target tracking.
Here, the target is the source of only one measurement per time-step.
One exemplary application is radar-based air surveillance.

However, for safe navigation in autonomous driving the shape or extent of a target cannot be neglected.
Moreover, with the upcoming of high-resolution LiDAR sensors, \eg the Velodyne HDL-64, a dense point cloud is available.
Generally, this implies several measurements per target.
In these cases, the small-target-assumption cannot be used anymore.
Therefore, at first, group target tracking approaches have been developed \cite{bena:overviewGT}.
Every target generates spatially distributed measurements around the target, where the group consists of multiple sub-objects, \eg edges or corners.
These sub-objects are tracked together as a group with common motion \cite{bena:Granstroen2016}.

This paper utilizes extended target tracking (ETT), where the target is the measurement source of a changing number of generally noisy and spatially distributed measurements \cite{bena:Granstroen2016}.
ETT is a nonlinear estimation problem, where the, potentially time varying, extent of the target has to be recursively computed \cite{bena:Granstroen2016}.
As mentioned in \cite{bena:Granstroen2016}, ETT is often falsely equated with contour tracking in computer vision, where the complete contour of the target can be extracted of a RGB image.
Contrarily, in ETT the target shape has to be estimated over time through a sparse set of measurements.

\begin{figure}[t!]
	\centering
	\includegraphics[width=\linewidth]{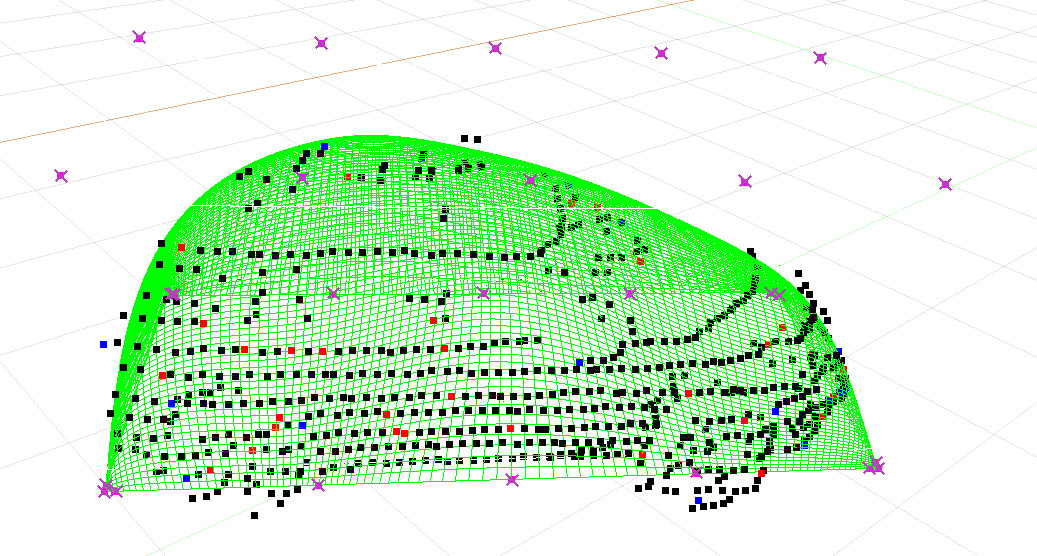}
	\caption{NURBS surface with estimated scaling parameters and weights. Crossed points denote the control points of the NURBS surface, whereas the blue and red points are used as measurements. The used measurements are sampled from the segmented point cloud of the vehicle.}\label{fig:result_nurbs}
\end{figure}

Common research in ETT deals with two-dimensional estimation in the x-y plane.
This paper operates with a rather dense 3D point cloud of the Velodyne-HDL64, where the 2D approximation does not hold.
Therefore, we propose a full three-dimensional estimation of the target's extent using the well-known NURBS surfaces \cite{bena:Piegl1991} as shape model.
NURBS surfaces originate from the geometric modeling community and are the building-block for accurate computer aided design models.
An exemplary result of estimating a NURBS surface with the proposed method is shown in \Cref{fig:result_nurbs}.

Our main contributions are:
\begin{itemize}
	\item A full 3D estimation of the target's extent while estimating the dynamics of the target
	\item A fast method where only the scaling of a pre-defined NURBS surface is estimated
	\item A second method which additionally estimates the weights of the NURBS surface, for more accurate shape estimation
	\item Both methods show real-time capability and are evaluated on real-world data against state-of-the-art
\end{itemize}

\section{Related Work} \label{sec:related_works}
An extensive overview of group target tracking is given by \cite{bena:overviewGT}, whereas \cite{bena:Granstroen2016} is a comprehensive ETT publication.
There exists a vast variety of shape models in ETT.
Prominent examples are sticks \cite{bena:Baum2012,bena:Gilholm2005} and rectangles \cite{bena:Granstroem2011}.
Moreover, a frequently used approximation is an ellipse.
There are several approaches for ellipse modeling like Gaussian inverse Wishart \cite{bena:lmbextended} or random matrix theory \cite{bena:Feldmann2011}.

Another class are the star-convex shapes.
The authors of \cite{bena:Baum2014} were one of the first to estimate arbitrary star-convex shapes by random hypersurface models (RHM).
In this work scaled versions of ellipses and circles as base representation of the RHM have been used.
The authors extended their work to arbitrary shapes by using level-set RHM \cite{LevelSetRHM} for general polygon trackers.
An additional method for star-convex shapes are Gaussian processes (GP), which have been utilized for many applications, \eg \cite{bena:wahlstromextended,bena:Martin2017}.
Up to now, only a small number of publications has dealt with the 3D case.
Recently, an extension for 3D GPs has been published \cite{bena:Kumru2018}.
Furthermore, \cite{bena:Faion2015} estimates the 3D objects by using transformed planes, whereas \cite{bena:Steinemann2012} estimates the 3D vehicle contour in LiDAR measurements.

B-Splines are a compact representation for a wide variety of shapes and they are, for instance, used for road estimation \cite{bena:Schreier2016,bena:burger_lane_2019_itsc}.
Additionally, there exist spline-based methods for tracking extended targets \cite{bena:Zea2016,bena:Daniyan2018,bena:Kaulbersch2018}.
\cite{bena:Kaulbersch2018} is the closest method to ours.
The authors model the target extent with a Cartesian 2D B-Spline curve.
For tracking vehicles, they estimate non-uniform scaling parameters in x- and y-dimension for static targets.
Contrarily, this paper uses a NURBS surface model to estimate a 3D target extent as well as the dynamics of the target.
Furthermore, \cite{bena:Kaulbersch2018} only evaluates on simulated data opposed to our real-world evaluations.

\section{Extended Target Tracking} \label{sec:problem}
To track an extended target, the standard state space model has to be augmented.
For instance, a shape model has to be specified to find the corresponding measurement source at the target's surface to the noise corrupted measurement.
Let $ \bm{x}_k^d $ be the additional kinematic parameters of a process model and $ \bm{x}_k^s $ be the shape part of the state.
Furthermore, let $ \bm{m}_k $ be the Cartesian center and $ \psi_k $ the orientation of the extended target.
Then, the augmented state vector is defined as follows:
\begin{equation}
\bm{x}_k = \begin{pmatrix} \bm{m}_k  &\psi_k & \bm{x}_k^d &  \bm{x}_k^s   \end{pmatrix}^T
\end{equation}
\subsection{Dynamic Model}
The following standard Gaussian dynamic model is chosen:
\begin{equation}
\bm{x}_{k} = f(\bm{x}_{k-1},\bm{v}_{k-1},\bm{u}_{k-1}),\;\;\bm{v}_{k-1} \sim \mathcal{N}(\bm{0},\bm{Q}_{k-1}),
\end{equation}
where $ \bm{Q}_{k-1} $ is the process noise covariance and $\bm{u}_{k-1}$ is the input. In general, the input is not known and therefore, it is approximated as Gaussian noise with $ \bm{u}_{k-1} \sim \mathcal{N}(\bm{0},\bm{C}_{k-1}) $.

In this work we applied the constant curvature and velocity (CCV) \cite{bena:Schuster2008} model as we deal with vehicles.
Hence, it follows $ \bm{x}_k^d = \begin{pmatrix} v_k & c_k \end{pmatrix} $ with velocity $ v_k $  and curvature $ c_k $.
Furthermore, $ \bm{u}_{k} $ is defined as $ \begin{pmatrix} \dot{v}_k & \dot{c}_k  \end{pmatrix}^T $.
\subsection{Shape Representation}
In general, an extended target is the source of a set of $ n_k\in \mathbb{N} $ noise corrupted measurements $ \bm{Y}_k =\lbrace \bm{y}_{kl} \rbrace_{l=0}^{n_k}$.
The relation between measurement source $ \bm{z}_{kl} $ and corrupted measurement is modeled as:
\begin{equation}
\bm{y}_{kl} = \bm{z}_{kl} + \bm{w}_{kl},\;\;\bm{w}_{kl} \sim \mathcal{N}(\bm{0},\bm{R}_{kl}),  \label{eq:noisy_meas}
\end{equation}
where $ \bm{R}_{kl} $ is the corresponding measurement noise covariance.
Associating $ \bm{y}_{kl} $ to its generating measurement source $ \bm{z}_{kl} $ is not a trivial task.
Therefore, the level-set of \cite{LevelSetRHM} as shape representation is chosen.
Let $ d(\bm{x}_k,\bm{z}_{kl}) $ be a shape function.
Then, the level-set for the level $ c \in \mathbb{R} $ is defined by \cite{LevelSetRHM}:
\begin{equation}
	\mathcal{L}_{d}(\bm{x}_k,c) =\left\lbrace \bm{z}_{kl} \mid d(\bm{x}_k,\bm{z}_{kl}) =c \right\rbrace.
\end{equation}
Clearly, the target shape boundary is given by $ \mathcal{L}_{d}(\bm{x}_k,0) $.
Moreover, the set of all $ \bm{z}_{kl} $ inside the shape is given by:
\begin{equation}
	\mathcal{S}_d(\bm{x}_k) =  \lbrace \bm{z}_{kl} \mid d(\bm{x}_k,\bm{z}_{kl} ) \geq 0\rbrace.
\end{equation}
\subsection{Measurement Model}
It is a challenging task to calculate and associate the corresponding level-set $ \mathcal{L}_{d}(\bm{x}_k,c) $ to every measurement source $ \bm{z}_{kl} $.
However, with Level-Set RHMs of \cite{LevelSetRHM} the explicit level-set does not have to be computed as Level-Set RHMs model it with a randomly distributed scaling of the maximum level.
Let $ \alpha_{kl} \sim \mathcal{U}(0,1)$ be an uniformly distributed scaling factor and $ d_{\max}(\bm{x_k}) = \max_{\hat{\bm{z}} \in \mathcal{S}_d(\bm{x}_k) } d(\bm{x}_k,\hat{\bm{z}})  $ be the upper bound of the shape function, which is the maximum level.
Then, it follows with \Cref{eq:noisy_meas} and $ \alpha_k \cdot  d_{\max}(\bm{x_k}) = d(\bm{x}_k,\bm{y}_{kl}-\bm{w}_{kl}) $
the measurement equation of one measurement \cite{LevelSetRHM}:
\begin{align}
0  &=  \alpha_{kl} \cdot d_{\max}(\bm{x}_k) - d(\bm{x}_k,\bm{y}_{kl}-\bm{w}_{kl})\\
&=g(\bm{x}_k,\bm{y}_{kl},\bm{w}_{kl},\alpha_{kl}).
\end{align}
Hence, the current state, scaling factor and measurement are mapped to the pseudo-measurement $ 0 $, which is the level for the shape boundary.

It is a common approach to model the likelihood of each measurement $ p(\bm{y}_{kl} \mid \bm{x}_k) $ conditionally independent, which is described by the following likelihood of all measurements:
\begin{equation}
p\left(\bm{Y}_k \mid \bm{x}_k \right) = \prod_{l=0}^{n_k} p(\bm{y}_{kl} \mid \bm{x}_k).
\end{equation}
This implies order independence of the incorporated measurements.
However, while sequentially updating the posterior of a Bayesian estimator with nonlinear measurements, the order of the measurements, potentially, changes the outcome \cite{LevelSetRHM,bena:Faion2012}.
Therefore, we apply all measurements at once with the following stacked measurement equation:
\begin{align}
\bm{g}\left(\bm{x}_k,\bm{Y}_{k},\bm{w}_k,\bm{\alpha}_k \right) &= \begin{pmatrix} g(\bm{x}_k,\bm{y}_{k0},\bm{w}_{k0},\alpha_{k0})^T\\\vdots\\g(\bm{x}_k,\bm{y}_{k n_k},\bm{w}_{kn_k},\alpha_{k n_k})^T  \end{pmatrix}, \\
\bm{R}_k &= diag\begin{pmatrix} \bm{R}_{k1},\dots,\bm{R}_{kn_k}  \end{pmatrix}, \\
 \bm{w}_k &\sim \mathcal{N}(\bm{0},\bm{R}_k),
\end{align} where $ \bm{\alpha}_k = \lbrace \alpha_{kl} \rbrace_{l=0}^{n_k} $.
Modeling $p\left(\bm{Y}_k \mid \bm{x}_k \right)$ with $ \bm{g}\left(\bm{x}_k,\bm{Y}_{k},\bm{w}_k,\bm{\alpha}_k\right) $ we get order independence.
\subsection{Inference}
Due to the high non-linearity of the measurement model as well as the presence of multiplicative noise, the Unscented Kalman Filter (UKF) \cite{UKFMerwe} is applied. The UKF is a sampling based Bayesian state estimator.

\section{NURBS Surface Model} \label{sec:filter}
In this part the used NURBS surface, the resulting NURBS shape function as well as the state's shape parts are explained.
\subsection{NURBS Surface Function}
\begin{figure}[t!]
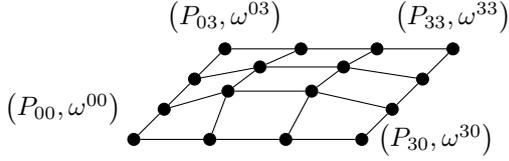

	\centering
	\include{nurbs}
	\caption{Exemplary control net of a NURBS surface with 16 control points $ P_{ij} $ and corresponding weights $ \omega^{ij} $, $ i=0,\dots,n_u$, $j=0,\dots,n_v $. Number of control points in $ u $-direction $  n_u $ and $ v $-direction $n_v $ equals 4.} \label{fig:nurbs}
\end{figure}
Let $P_{ij} \in \mathbb{R}^{3 \times (n_u + 1) \times (n_v + 1)}  $ be the bidirectional control net tensor and $ n_u,n_v $ the number of control points in $ u $- and $ v $-direction. Furthermore, let $\lbrace N_{ip}(u) \rbrace_{i=0}^{n_u},\lbrace N_{jq}(v) \rbrace_{j=0}^{n_v}$ be the $ p$-th and $ q $-th degree B-spline basis functions, which are defined on the knot sequences $ U=\lbrace u_l\rbrace_{l=0}^{n_u + p+ 1} $ and $ V=\lbrace v_l\rbrace_{l=0}^{n_v +q+ 1} $.
Moreover, let $ \bm{s}_k = \begin{pmatrix}s_k^x&s_k^y & s_k^z \end{pmatrix}^T $ with $ s_k^x,s_k^y, s_k^z \in \mathbb{R}_{>0}$ be scaling factors on every axis and $\bm{\omega}_k = \lbrace \omega_k^{ij} \rbrace_{i,j=0}^{n_u \times n_v} $ the weights of control points.
Then, our modified NURBS surface function is defined with the following tensor product scheme \cite{bena:Piegl1991}:
\begin{equation}
\bm{S}_k(u,v,\bm{\omega}_k,\bm{s}_k) = \bm{s}_k \bullet \frac{\sum\limits_{i=0}^{n_u} \sum\limits_{j=0}^{n_v}N_{ip}(u)N_{jq}(v)\omega_k^{ij}P_{ij}}{\sum\limits_{i=0}^{n_u} \sum\limits_{j=0}^{n_v}N_{ip}(u)N_{jq}(v)\omega_k^{ij}},
\end{equation}
with  $0 \leq u,v\leq 1$ and $ (\bullet) $ is the Hadamard product.
\Cref{fig:nurbs} illustrates a control net of a NURBS surface.
For better readability we set $\bm{S}_k(u,v) := \bm{S}_k(u,v,\bm{\omega}_k,\bm{s}_k)$.

The measurement model is defined in target coordinates.
Therefore, the measurement source in target coordinates $ \hat{\bm{z}}_{kl} $ is defined as:
\begin{equation}
\hat{\bm{z}}_{kl} = R_{\psi_k}^{-1}\left(\bm{z}_{kl} - \bm{m}_k\right)
\end{equation}
Now, the two proposed methods are introduced.
Method 1 (M1) has a better shape approximation capability, whereas Method 2 (M2) is significantly faster.
\subsubsection{Method 1 (M1)}
Here, we estimate scaling factors as well as the weights of the NURBS surface.
Consequently, the shape-part of the state is:
\begin{equation}
\bm{x}^s_k = \begin{pmatrix}s_k^x&s_k^y & s_k^z& \omega_k^{00}&\omega_k^{01} &... & \omega_k^{n_u  n_v} \end{pmatrix}.
\end{equation}
For NURBS surface smoothing, the weights have to be regularized.
An important property for the smoothness of a surface is the Gaussian curvature.
Therefore, we regularize the weights according to their corresponding Gaussian curvature in similar fashion to \cite{bena:Saini2015}.
Let $ \bm{I}(u,v) $, and $ \bm{II}(u,v) $ be the first and second fundamental form of a surface at a point $ (u,v) $.
Furthermore, set for better readability $ \bm{I}:=\bm{I}(u,v)  $, $ \bm{II}:=\bm{II}(u,v)$ and $ \bm{S}_k:=\bm{S}_k(u,v) $.
Then, $ \bm{I} $ and $ \bm{II} $ are defined as \cite{bena:Goldman2005}:
\begin{align}
   \bm{I} &= \begin{pmatrix} \left\langle \bm{S}_k^{(u)},\bm{S}_k^{(u)}\right\rangle & \left\langle  \bm{S}_k^{(u)},\bm{S}_k^{(v)}\right\rangle\\
   \left\langle \bm{S}_k^{(v)},\bm{S}_k^{(u)}\right\rangle & \left\langle \bm{S}_k^{(v)},\bm{S}_k^{(v)}\right\rangle\end{pmatrix} \\
   \bm{II} &= \begin{pmatrix} \left\langle \bm{S}_k^{(uu)},\bm{N}_k\right\rangle & \left\langle \bm{S}_k^{(uv)},\bm{N}_k\right\rangle\\
   \left\langle \bm{S}_k^{(vu)},\bm{N}_k\right\rangle & \left\langle \bm{S}_k^{(vv)},\bm{N}_k\right\rangle\end{pmatrix},
\end{align}
where $ \bm{N}_k = \nicefrac{\bm{S}_k^{(u)} \times \bm{S}_k^{(v)}}{\|\bm{S}_k^{(u)}\times \bm{S}_k^{(v)}\|_2} $ is the normal vector and exemplary $ \bm{S}_k^{(u)} $ the partial derivative with respect to $ u $.
It follows the Gaussian curvature $ K_G(u,v) $ of the surface patch corresponding to $ (u,v) $ with \cite{bena:Goldman2005}:
\begin{equation}
	K_G(u,v) = \frac{|\bm{II}|}{|\bm{I}|},
\end{equation}where $ |B| $ denotes the determinant of a matrix $ B $.
Hence, the dynamic model for one of the state's weights is defined as follows:
\begin{equation}
   \omega^{ij}_{k} =    \omega^{ij}_{k-1}+ \nu \cdot \frac{ K_G(u,v) }{\max_{(\tilde{u},\tilde{v})\in U\times V}K_G(\tilde{u},\tilde{v})} + v_{k-1}^{\omega^{ij}},
\end{equation} where the Gaussian curvature at the surface point is normalized through the maximum Gaussian curvature of all surface points. Furthermore, $ v_k^{\omega^{ij}} $ is the additive Gaussian process noise of weight $ w_k^{ij} $ and $ \nu $ is a damping factor.
\subsubsection{Method 2 (M2)}
In this method, only the scaling factors are estimated.
Therefore, set $\omega_k^{ij} = 1$, $ \forall k $ and $ i=0,\dots,n_u $, $ j=0,\dots,n_v $.
Then, the reduced shape-part of the state is:
\begin{equation}
\bm{x}^s_k = \begin{pmatrix}s_k^x&s_k^y & s_k^z \end{pmatrix}.
\end{equation}
\subsection{NURBS Shape Function}
In this section we define the shape function, which is based on the NURBS surface and used in the measurement model.
But first, the closest surface point $ \bm{S}_k(\hat{u},\hat{v}) $ to a measurement source in local coordinates $ \hat{\bm{z}}_{kl} $ has to be found.
The chosen criterion is the angle difference $\angle \left(\bm{S}_k(u,v),\hat{\bm{z}}_{kl} \right) $, as we deal with dynamic and star-convex targets like cars.
For non-convex targets this could be easily extended with the minimum distance criterion of \cite{LevelSetRHM,bena:Kaulbersch2018}.
Therefore, the parameter tuple $ (\hat{u},\hat{v}) $ corresponding for the surface point $ \bm{S}_k(\hat{u},\hat{v}) $ is calculated through:
\begin{align}
(\hat{u},\hat{v}) &= \argmin\limits_{(u,v) \in U \times V } \angle \left(\bm{S}_k(u,v),\hat{\bm{z}}_{kl} \right).
\end{align}

Another premise of the shape function is the mahalonobis distance $ m(\bm{x}_k,\hat{\bm{z}}_{kl} ) $, which is defined as:
\begin{equation}
m(\bm{x}_k,\hat{\bm{z}}_{kl}) = \sqrt{ \left( \hat{\bm{z}}_{kl} -\bm{S}_k(\hat{u},\hat{v})\right)^T \cdot \bm{R}_{kl}^{-1} \cdot \left(\hat{\bm{z}}_{kl} - \bm{S}_k(\hat{u},\hat{v}) \right) }.
\end{equation}
It follows the NURBS shape function as the signed mahalonobis distance:
\begin{equation}
d(\bm{x}_k,\hat{\bm{z}}_{kl}) = \begin{cases}
m(\bm{x}_k,\hat{\bm{z}}_{kl}), &\mathtt{if}\;\; \hat{\bm{z}}_{kl} \in \mathcal{S}_d(\bm{x}_k) \\
- m(\bm{x}_k,\hat{\bm{z}}_{kl}), &\mathtt{else} \\
\end{cases}.
\end{equation}

\section{Results} \label{sec:results}
The proposed methods are evaluated in real-world scenarios.
They are recorded with a roof-mounted Velodyne HDL64-S2 at our institute's autonomous car.
Ground truth data is obtained through an equipped inertial navigation system (INS) sensor at the target.
Video footage can be found online\footnote{https://youtu.be/1tL3UrLhAUE}.
Performance evaluations are done against state-of-the-art methods, as 3D Gaussian-processes (3DGP) \cite{bena:Kumru2018} and the 2D Cartesian B-Spline method of \cite{bena:Kaulbersch2018} (2DBS).
Furthermore, we compare against a simple point tracking approach (PT) \cite{bena:GDPF}, where the bounding box center as well as their dimensions are estimated.
The segmented point clouds of the target vehicle are provided from the method of \cite{bena:burger_iv2018,bena:burger_itsc2018}.
During the scenario, different occlusions (only the back or one side can be seen) of the segmented vehicle and segmentation errors, \eg falsely associated ground plane points, occur.
Moreover, the bounding boxes of the PT approach are obtained with \cite{bena:BeNaIV2018}.
The chosen number of points in the scenarios are a compromise between better estimation results and faster run time.
\begin{figure*}[ht!]
	\centering
	\begin{subfigure}{0.24\linewidth}
		\centering\includegraphics[trim={150 200 150 200},clip,width=0.9\linewidth]{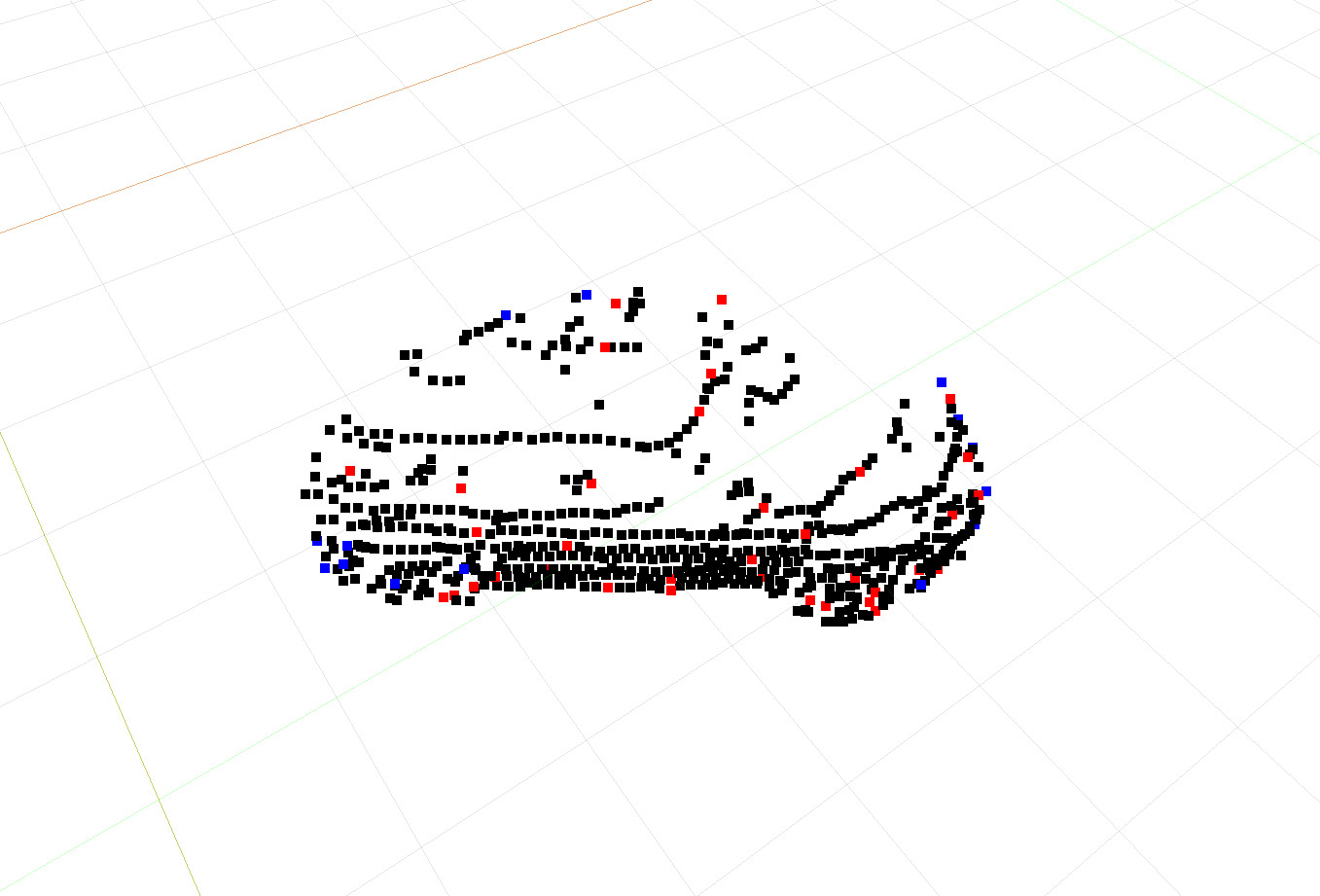}
		\caption{Segmented point cloud and sampled points in red and blue.} \label{fig:car_points}
	\end{subfigure}
	\begin{subfigure}{0.24\linewidth}
		\centering\includegraphics[trim={150 200 150 200},clip,width=0.9\linewidth]{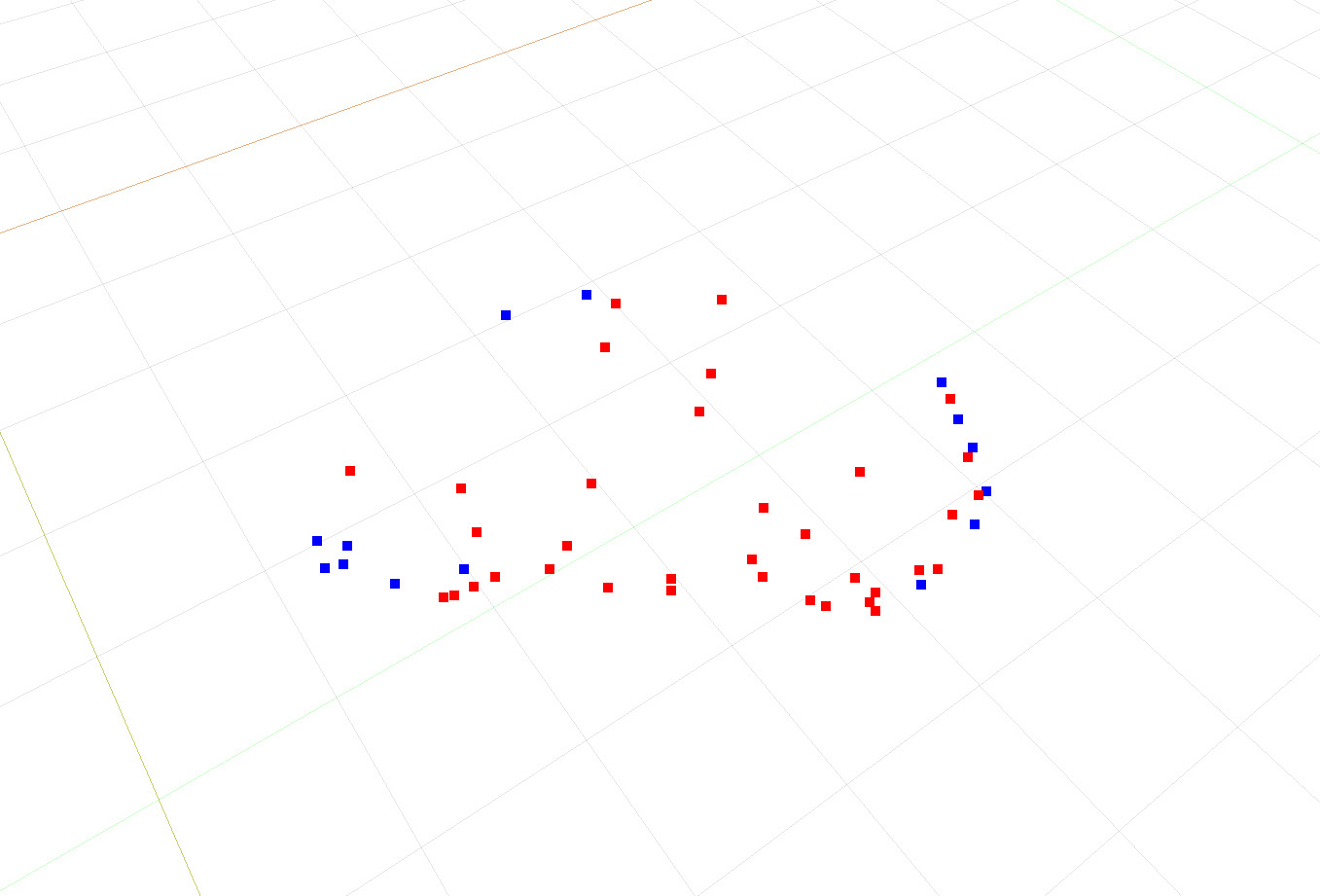}
		\caption{50 sampled points used as measurements in the static scenario.}\label{fig:car_points_50}
	\end{subfigure}
	\begin{subfigure}{0.24\linewidth}
	\centering\includegraphics[trim={150 200 150 200},clip,width=0.9\linewidth]{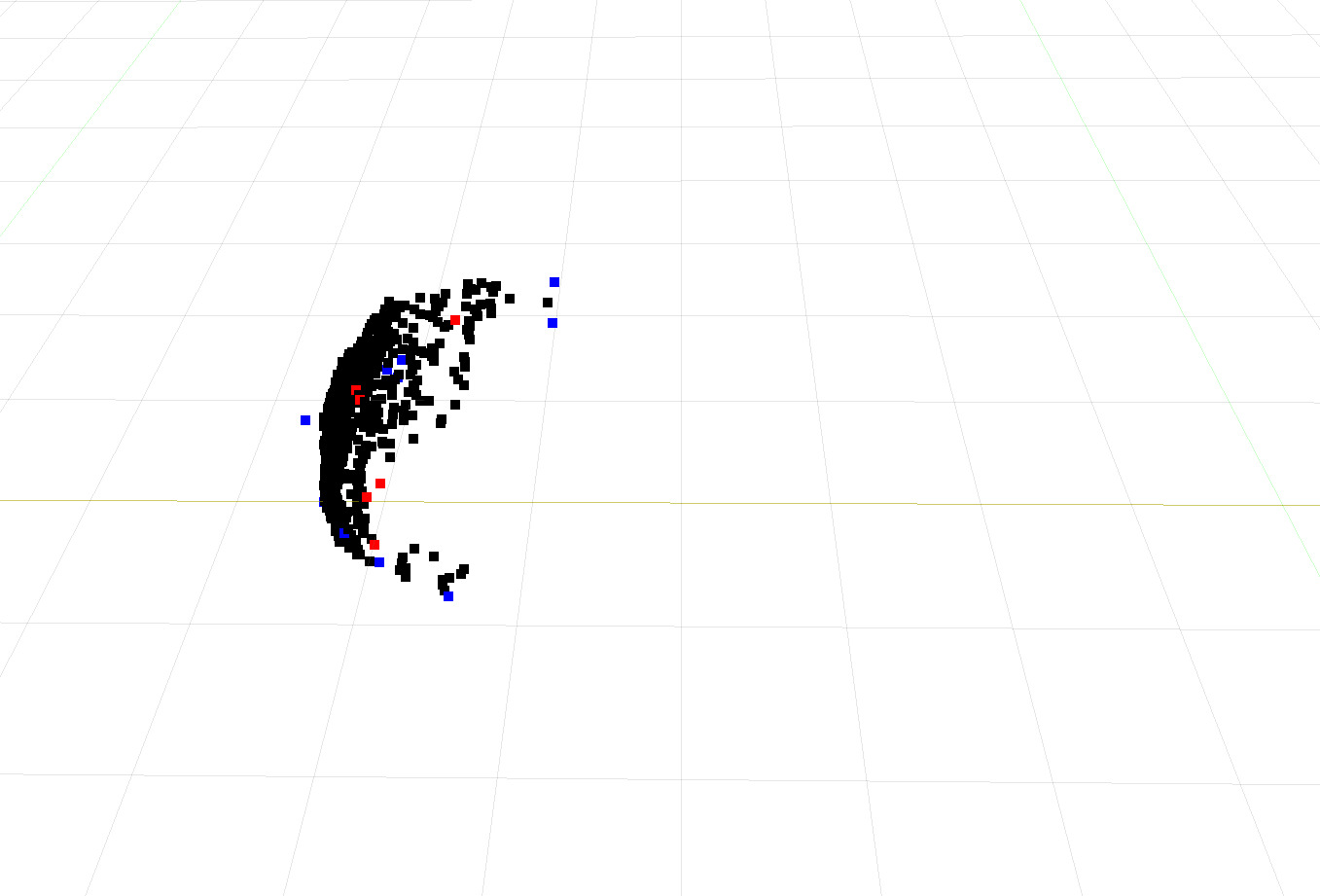}
	\caption{Segmentation of occluded vehicle with sampled points in red and blue.}\label{fig:car_points_back}
	\end{subfigure}
	\begin{subfigure}{0.24\linewidth}
	\centering\includegraphics[trim={150 200 150 200},clip,width=0.9\linewidth]{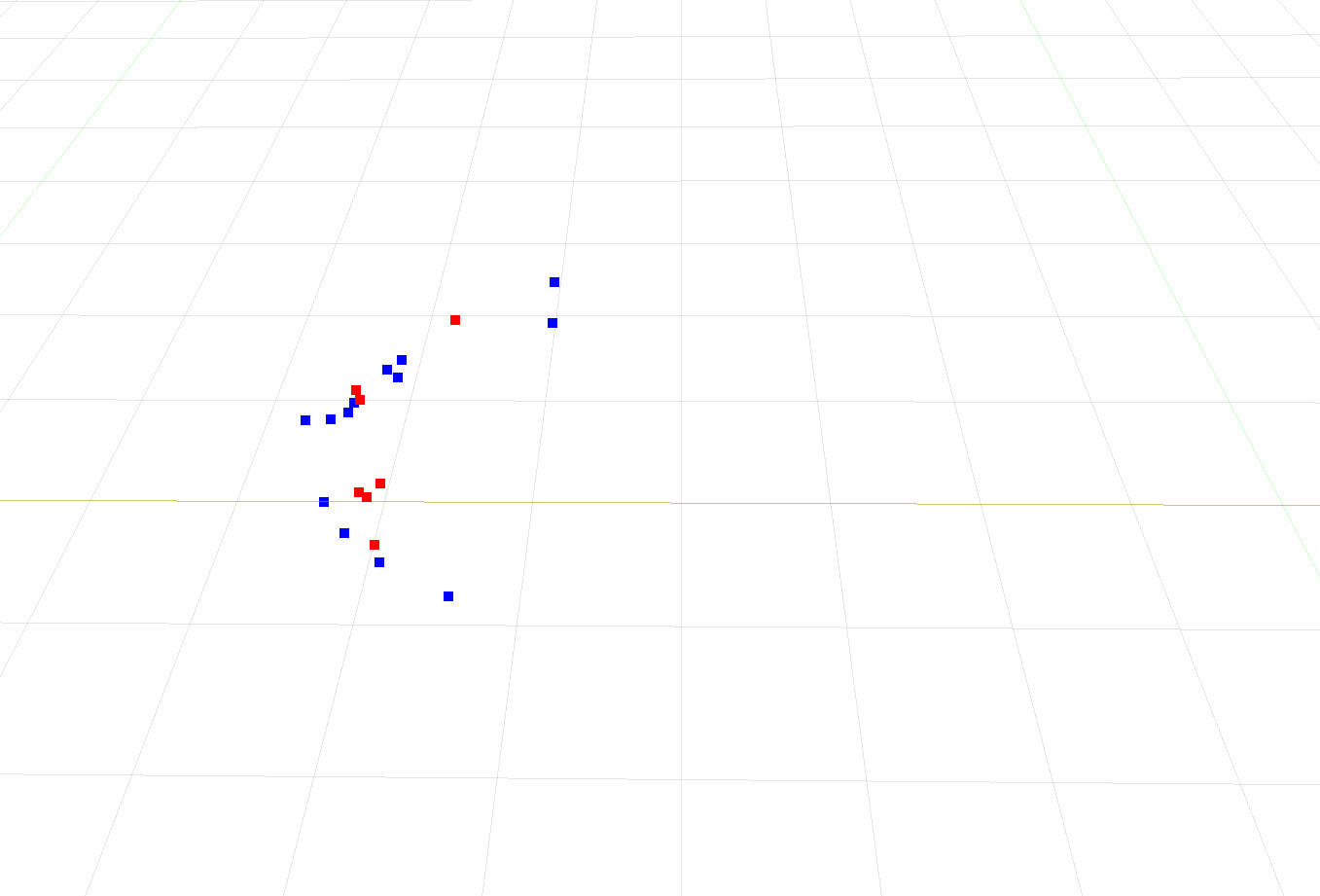}
	\caption{20 points used as measurements in the dynamic scenario.}\label{fig:car_points_back_20}
\end{subfigure}
	\caption{Illustration of different segmentation of a vehicle. The blue color correspond to points originating from the 2D convex hull, whereas red points are randomly sampled over the whole point cloud.}
\end{figure*}
\begin{figure*}[ht!]
	\centering
	\begin{subfigure}{.24\linewidth}
		\centering\includegraphics[trim={0 200 0 100},clip,width=0.9\linewidth]{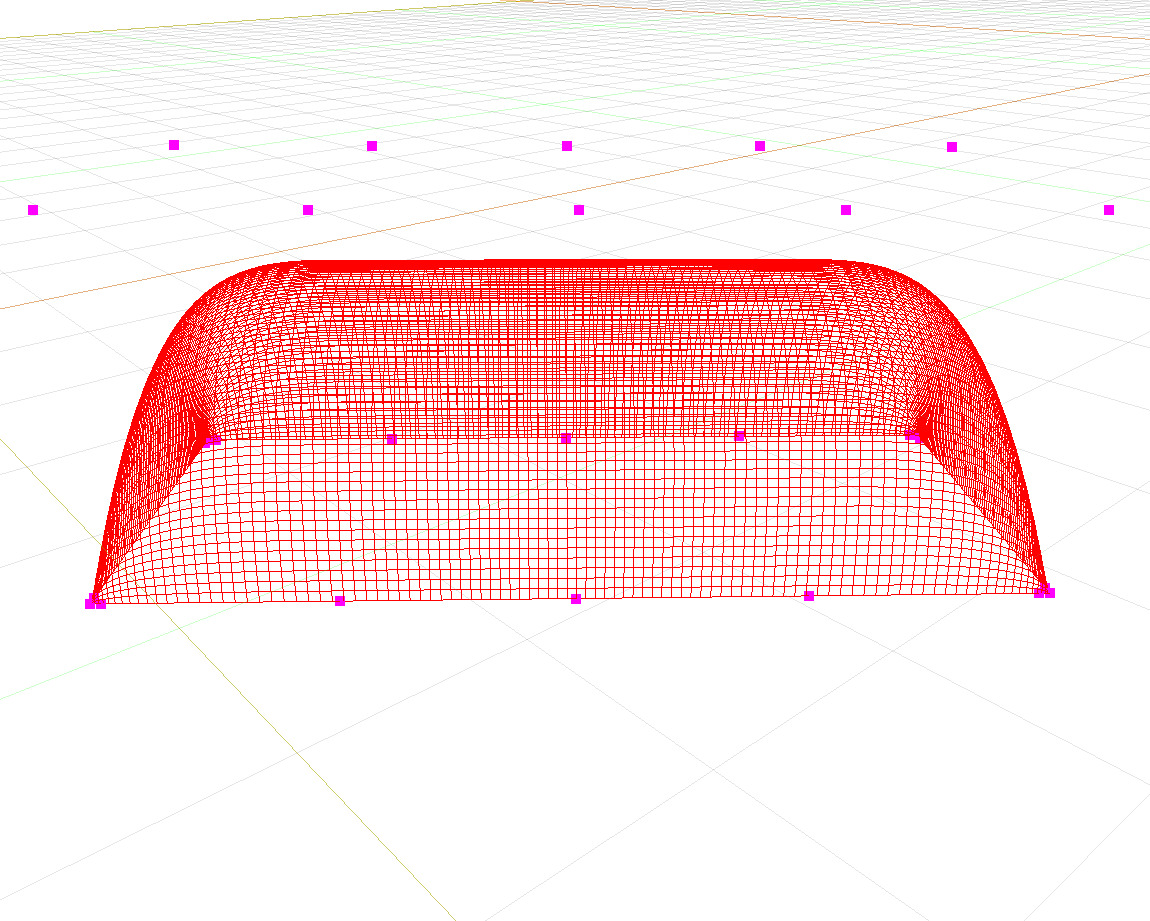}\\
		\includegraphics[trim={0 200 0 100},clip,width=0.9\linewidth]{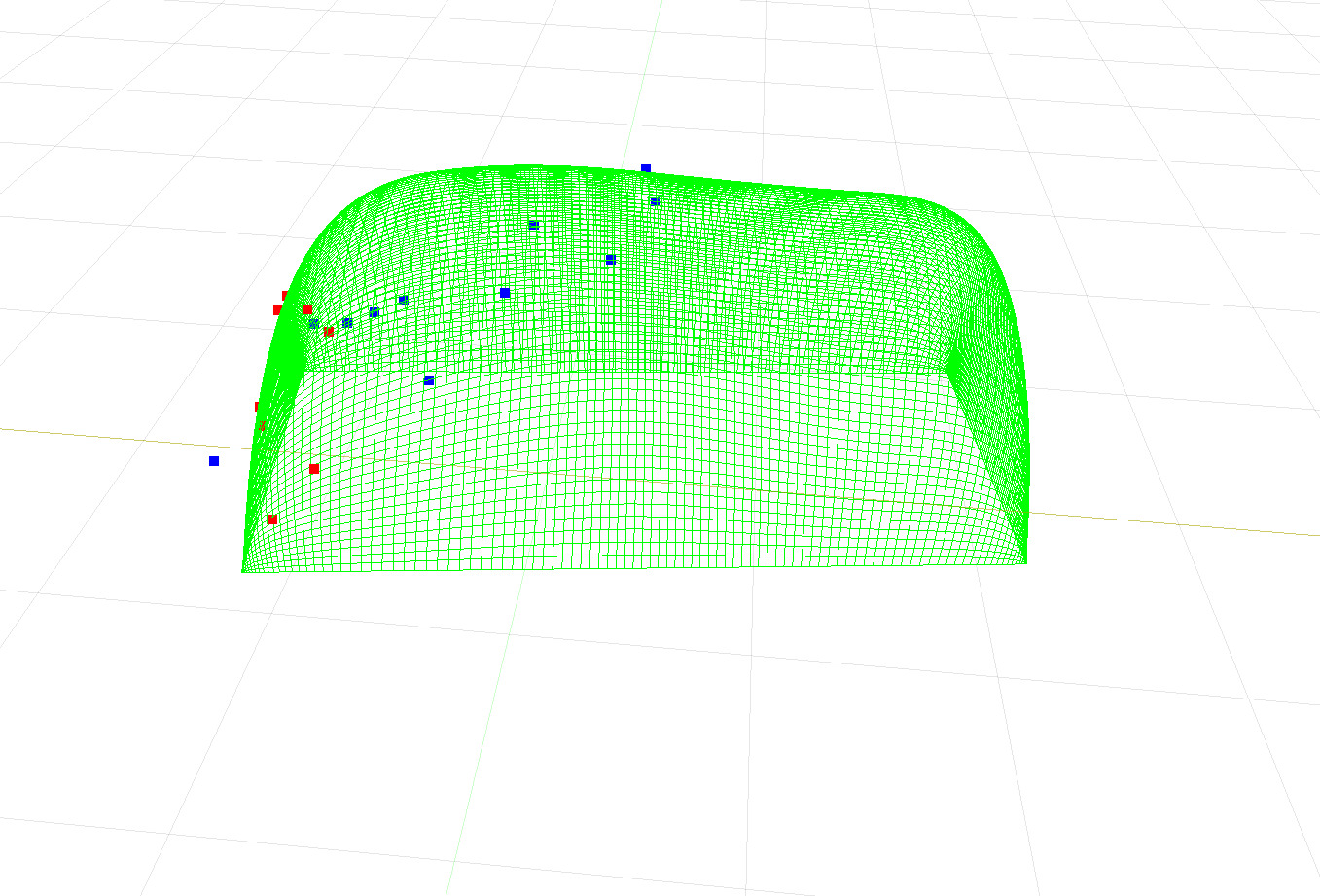}
		\caption{M1 with 28 control points.} \label{fig:init_nurbs}
	\end{subfigure}
	\begin{subfigure}{.24\linewidth}
		\centering\includegraphics[trim={0 200 0 100},clip,width=0.9\linewidth]{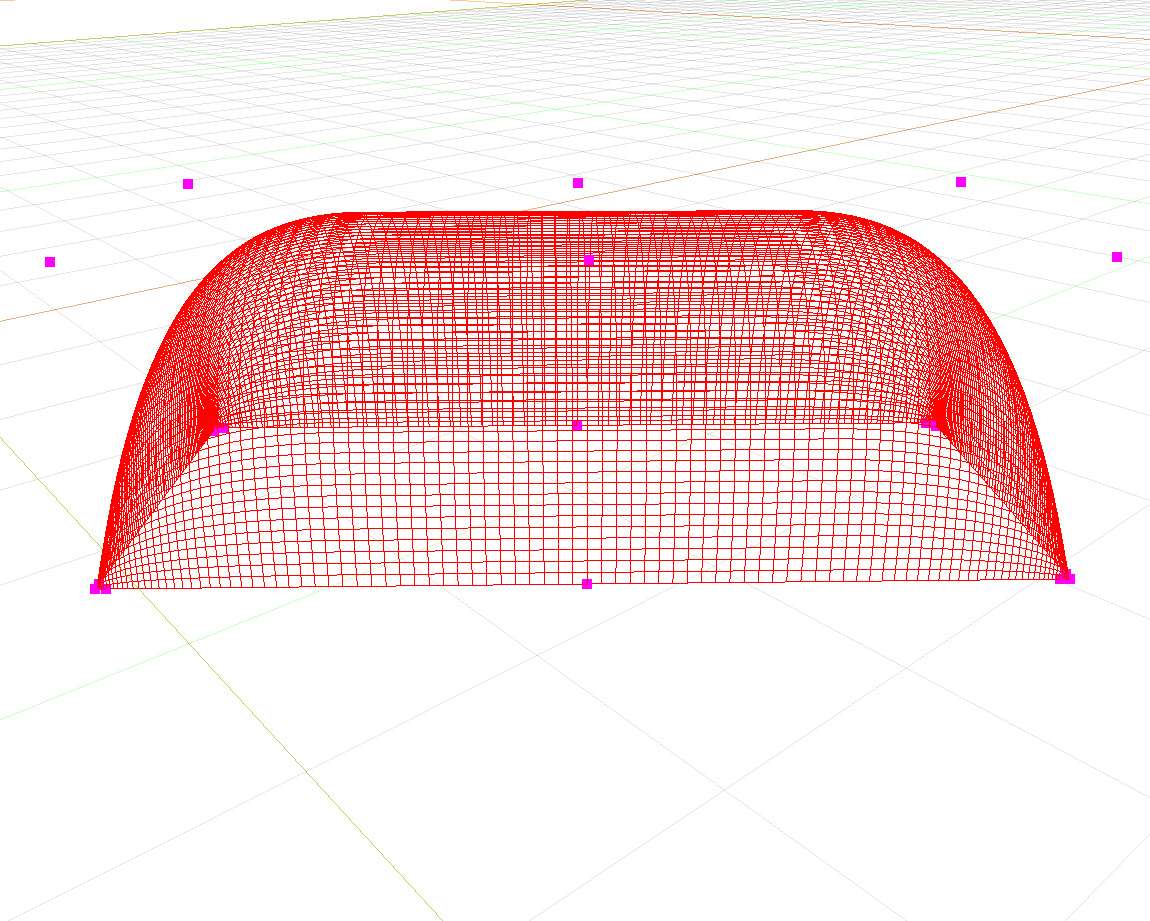}\\
		\includegraphics[trim={0 200 0 150},clip,width=0.9\linewidth]{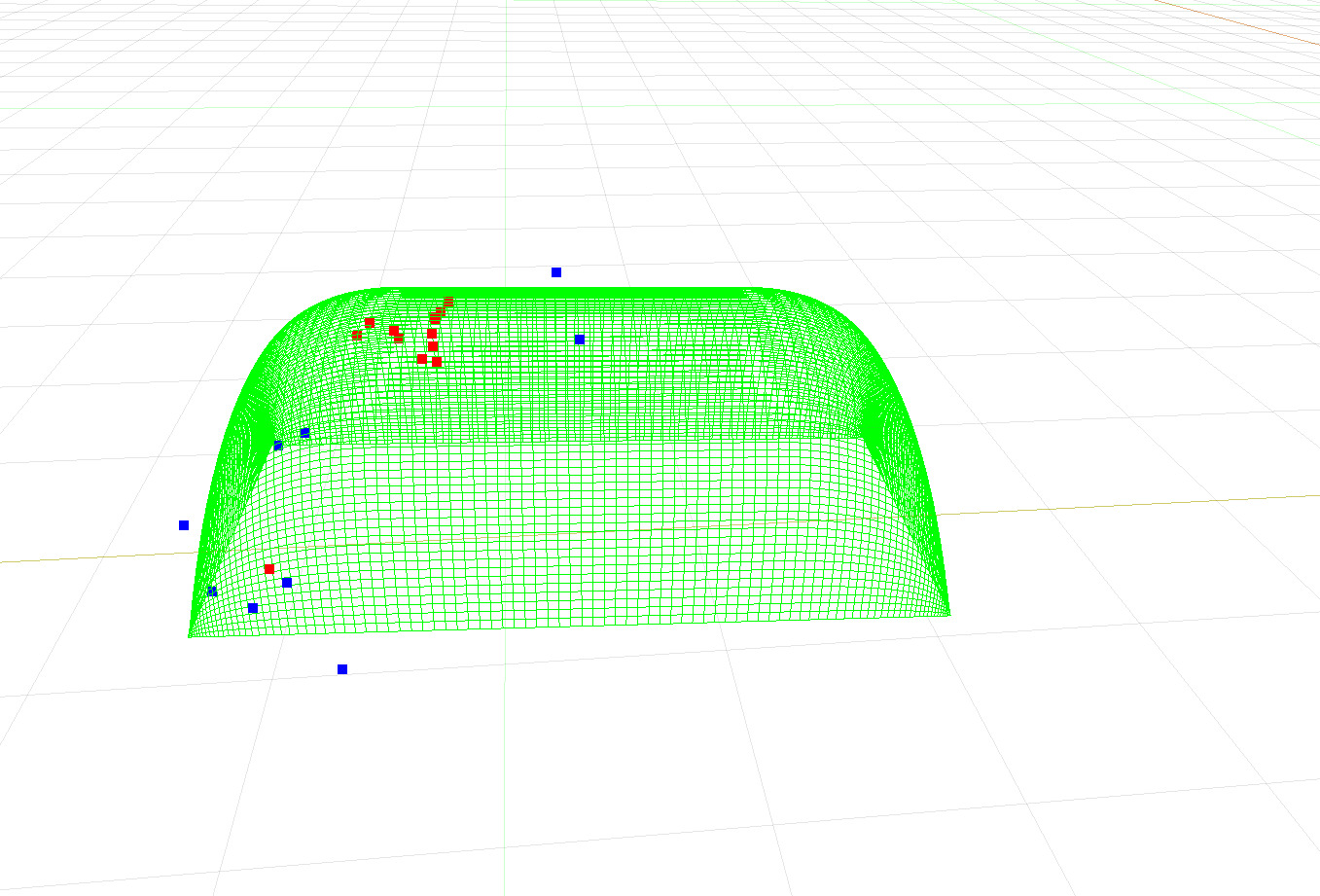}
		\caption{M2 with 20 control points.}\label{fig:init_surface}
	\end{subfigure}
	\begin{subfigure}{.24\linewidth}
		\centering\includegraphics[trim={0 200 0 150},clip,width=0.9\linewidth]{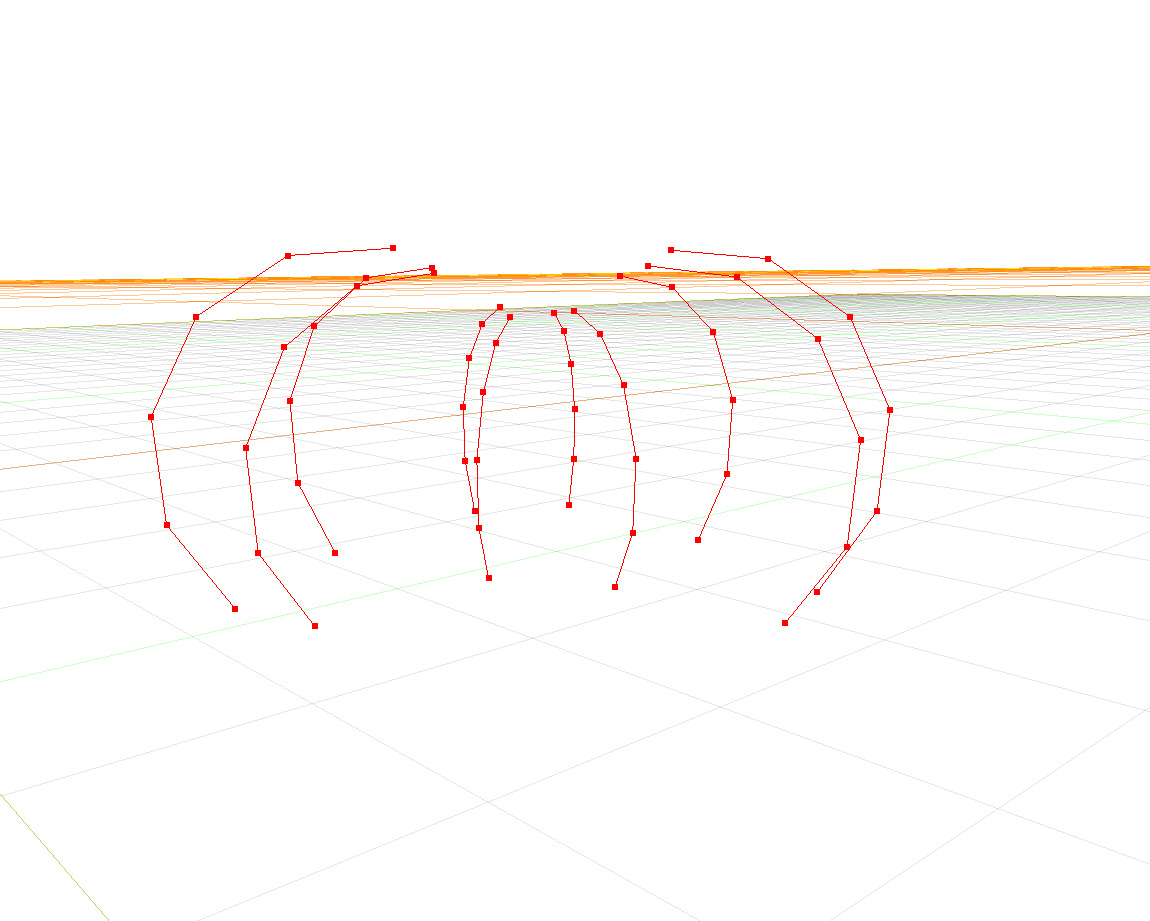}
		\includegraphics[trim={0 200 0 150},clip,width=0.9\linewidth]{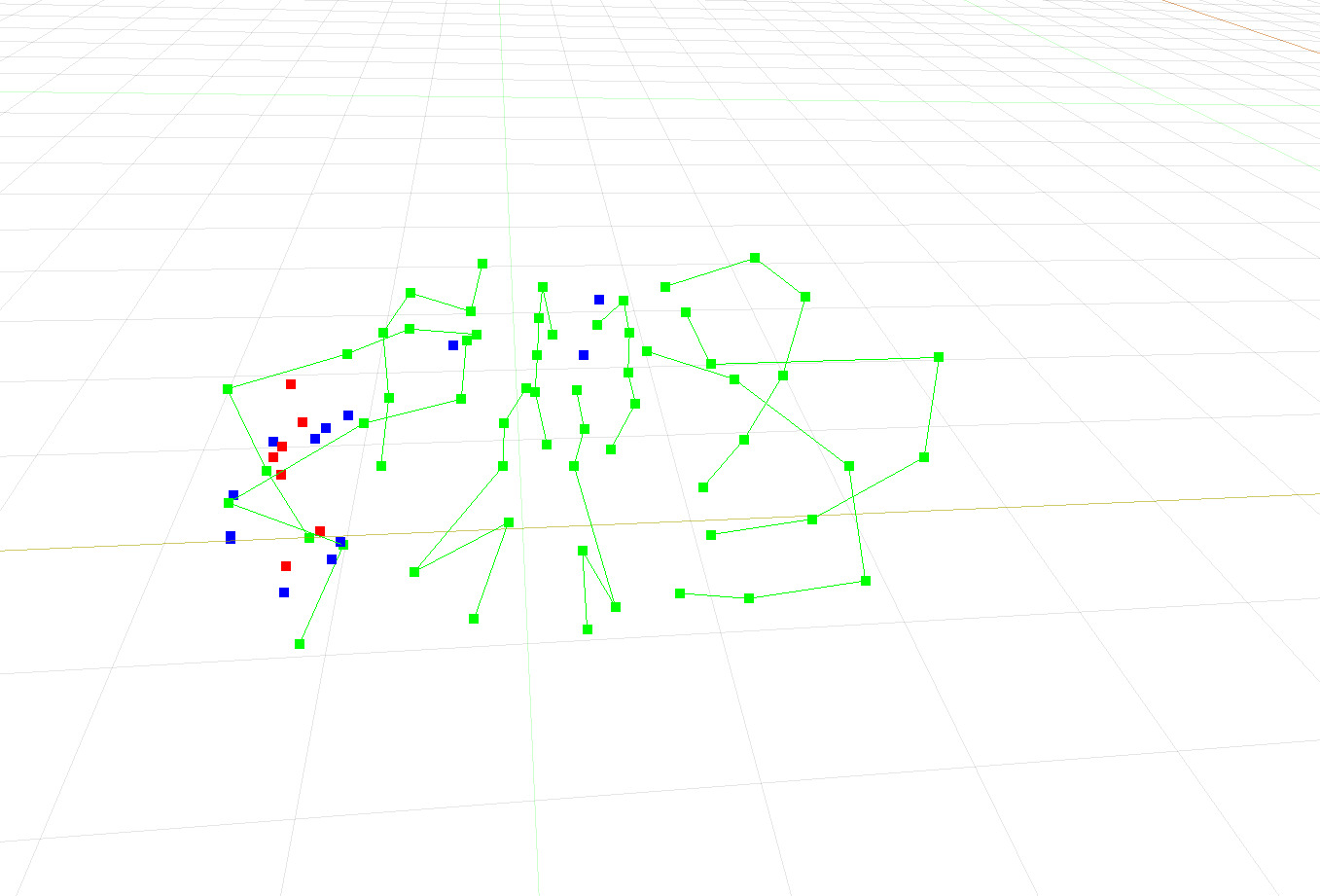}
		\caption{3DGP with 60 control points.}\label{fig:init_gp}
	\end{subfigure}
	\begin{subfigure}{.24\linewidth}
		\centering\includegraphics[trim={0 200 0 100},clip,width=0.9\linewidth]{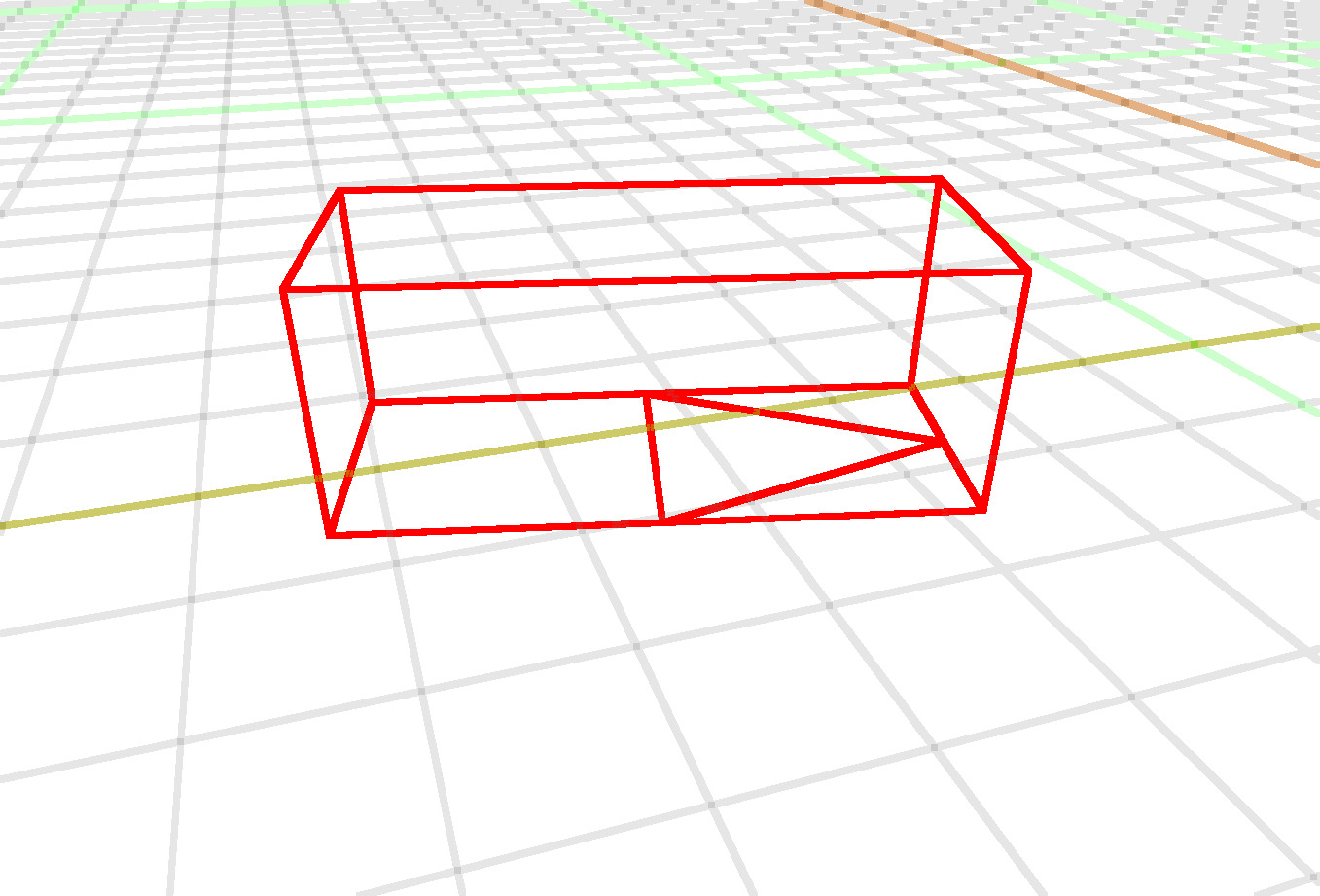}\\
		\includegraphics[trim={0 200 0 120},clip,width=0.9\linewidth]{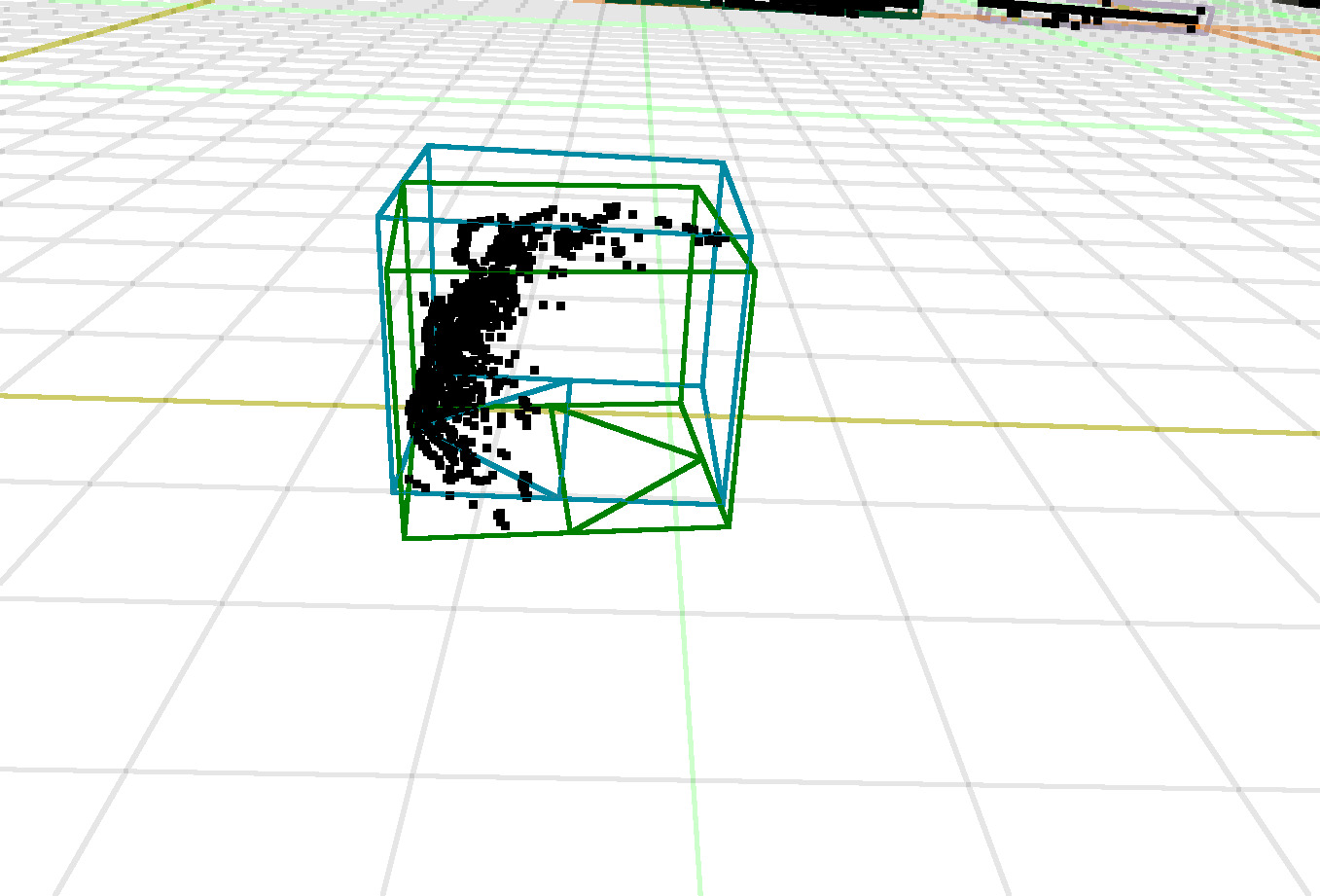}
		\caption{SP with bounding box.}\label{fig:init_point}
	\end{subfigure}
	\caption{The initial shape of the methods is shown in the upper half, where the estimated shape is shown in the lower half. The measurements (red and blue points) only originate from the back of the vehicle.} \label{fig:quali_dynamic}
\end{figure*}
\subsection{Scenario 1: Static Object}
The goal of this scenario is to estimate the correct shape of a static vehicle.
Therefore, the static vehicle is recorded from every side by driving around it multiple times.
As measurements, we randomly sample 50 laser points of the segmented object and its 2D convex hull.
\Cref{fig:car_points,fig:car_points_50} illustrate the object's point cloud sampling.
\subsubsection{Initialization}
For M1, we initialize all weights $ \bm{\omega}_0 $ and scaling factors $ \bm{s}_0 $ to $ 1 $.
The damping factor for regularization is set to $ \nu = 0.001$.
Moreover, shape covariances are set to $ Q_{w^{ij}} = 0.1 \;\forall i,j$ and $ Q_{s^x} = Q_{s^y} = Q_{s^z} = 10^{-7}$.
Furthermore, we use $ n_u=7 $, $ n_v=4 $ with cubic basis splines, which means: $ p=q=3 $.
The red surface in \Cref{fig:init_nurbs} shows the initial configuration of M1's NURBS surface.

M2 has the same initial scaling factors, and shape covariances for the scaling factors as M1.
However, we use quadratic basis splines ($ p=q=2 $) and less control points ($ n_u=5  $, $ n_v=4 $).
The initial surface of M2 can be seen in the upper half of \Cref{fig:init_surface}.
3DGP is set up with 60 control points, which can be seen in the upper side of \Cref{fig:init_gp}.
The hyper parameters of 3DGP are the variance of the mean value $ \sigma_r $, the prior variance $ \sigma_f $ and the length scale $ l_{gp} $, which are set to $  \sigma_r=0.5 $, $ \sigma_f= 2 $ and $ l_{gp}=\nicefrac{ \pi}{20} $.
2DBS is initialized with 20 control points as in the original paper.
We have tried the method with different parameter settings.
However, 2DBS seems to be unable to estimate 3D point data as it does not cope for measurements inside the closed B-spline curve.
The simple point tracking approach is initialized with the first detected bounding box, which is shown in \Cref{fig:init_point} as the red bounding box.

Covariances for input noise are set to $ \bm{C}_{\dot{v}} = 10^{-4} $ and $ \bm{C}_{\dot{c}} = 10^{-4} $. 
The input noise for velocity and curvature is set to small values as we do not expect movements of the target.
\subsubsection{Evaluation}
\begin{figure*}[t!]
	\centering
	\begin{tikzpicture}
\begin{axis}[
height=2.7cm,
width=\linewidth,
legend style={legend cell align=left, align=right, draw=white!15!black,nodes={scale=0.55, transform shape}},
legend pos = south east,
xlabel style={at={(0.999,0.4)},anchor=north east},
axis x line*=bottom,
axis y line*=left,
axis line style={->},
xmin=0,
y tick label style={
	/pgf/number format/.cd,
	fixed,
	fixed zerofill,
	precision=2,
	/tikz/.cd
},
xlabel=$ k $,
y unit=m/s,
ylabel=$ v_k$,
]
\addplot[color=red , line width=1pt] table[y expr={abs(\thisrow{gv})},x expr=\coordindex,col sep=comma]{nurbs_result.csv};
\addplot[color=blue , line width=1pt] table[y expr={abs(\thisrow{gv})},x expr=\coordindex,col sep=comma]{surface_result.csv};
\addplot[color=green , line width=1pt] table[y expr={abs(\thisrow{gv})},x expr=\coordindex,col sep=comma]{gp_result.csv};
\addplot[color=black , line width=1pt] table[y expr={abs(\thisrow{gv})},x expr=\coordindex,col sep=comma]{point_result.csv};
\legend{M1,M2,3DGP,SP}
\end{axis}
\end{tikzpicture}
\\
\begin{tikzpicture}
\begin{axis}[
height=3cm,
width=\linewidth,
legend style={legend cell align=left, align=right, draw=white!15!black,nodes={scale=0.55, transform shape}},
legend pos = south east,
xlabel style={at={(0.999,0.4)},anchor=north east},
axis x line*=bottom,
axis y line*=left,
axis line style={->},
y tick label style={
	/pgf/number format/.cd,
	fixed,
	fixed zerofill,
	precision=2,
	/tikz/.cd
},
xmin=0,
xlabel=$ k $,
y unit=m^2,
ylabel=$ e_A$,
]
\addplot[color=blue , line width=1pt] table[y expr={abs((\thisrow{l}*\thisrow{w})-(\thisrow{insl}*\thisrow{insw})},x expr=\coordindex,col sep=comma]{surface_result.csv};
\addplot[color=red , line width=1pt] table[y expr={abs((\thisrow{l}*\thisrow{w})-(\thisrow{insl}*\thisrow{insw})},x expr=\coordindex,col sep=comma]{nurbs_result.csv};
\addplot[color=green , line width=1pt] table[y expr={abs((\thisrow{l}*\thisrow{w})-(\thisrow{insl}*\thisrow{insw})},x expr=\coordindex,col sep=comma]{gp_result.csv};
\addplot[color=black , line width=1pt] table[y expr={abs((\thisrow{l}*\thisrow{w})-(\thisrow{insl}*\thisrow{insw})},x expr=\coordindex,col sep=comma]{point_result.csv};
\end{axis}
\end{tikzpicture}
\\
\begin{tikzpicture}
\begin{axis}[
height=2.4cm,
width=\linewidth,
legend style={legend cell align=left, align=left, draw=white!15!black},
legend pos = south west,
xlabel style={at={(0.999,0.4)},anchor=north east},
axis x line*=bottom,
axis y line*=left,
axis line style={->},
y tick label style={
	/pgf/number format/.cd,
	fixed,
	fixed zerofill,
	precision=2,
	/tikz/.cd
},
xmin=0,
xlabel=$ k $,
y unit=m,
ylabel=$ e_{pos}$,
]
\addplot[color=blue, line width=1pt] table[y expr={abs(\thisrow{ginssx}-\thisrow{gx})+abs(\thisrow{ginssy}-\thisrow{gy}))},x expr=\coordindex,col sep=comma]{surface_result.csv};
\addplot[color=red, line width=1pt] table[y expr={abs(\thisrow{ginssx}-\thisrow{gx})+abs(\thisrow{ginssy}-\thisrow{gy}))},x expr=\coordindex,col sep=comma]{nurbs_result.csv};
\addplot[color=black, line width=1pt] table[y expr={abs(\thisrow{ginssx}-\thisrow{gx})+abs(\thisrow{ginssy}-\thisrow{gy}))},x expr=\coordindex,col sep=comma]{point_result.csv};

\addplot[color=green , line width=1pt] table[y expr={abs(\thisrow{ginssx}-\thisrow{gx})+abs(\thisrow{ginssy}-\thisrow{gy}))},x expr=\coordindex,col sep=comma]{gp_result.csv};
\end{axis}
\end{tikzpicture}
\\
\begin{tikzpicture}
\begin{axis}[
height=2.4cm,
width=\linewidth,
legend style={legend cell align=left, align=left, draw=white!15!black},
legend pos = south west,
xlabel style={at={(0.999,0.4)},anchor=north east},
axis x line*=bottom,
axis y line*=left,
axis line style={->},
y tick label style={
	/pgf/number format/.cd,
	fixed,
	fixed zerofill,
	precision=2,
	/tikz/.cd
},
xmin=0,
xlabel=$ k $,
y unit=rad,
ylabel=$ e_{\psi_k}  $,
]
\addplot[color=blue , line width=1pt] table[y expr={abs(\thisrow{gyaw}-\thisrow{ginsyaw}},x expr=\coordindex,col sep=comma]{surface_result.csv};
\addplot[color=red , line width=1pt] table[y expr={abs(\thisrow{gyaw}-\thisrow{ginsyaw}},x expr=\coordindex,col sep=comma]{nurbs_result.csv};
\addplot[color=green , line width=1pt] table[y expr={abs(\thisrow{gyaw}-\thisrow{ginsyaw}},x expr=\coordindex,col sep=comma]{gp_result.csv};

\end{axis}
\end{tikzpicture}
	\caption{Results of estimating a static vehicle, with velocity $ v_k $, area error $ e_A $, Cartesian position error $ e_{pos} $ and orientation error $ e_\psi $. The area of the estimated shapes is defined as the product of the length and width of the encasing rectangle. Furthermore, one time step $ k $ is \SI{100}{\milli\second}.} \label{fig:static}
\end{figure*}
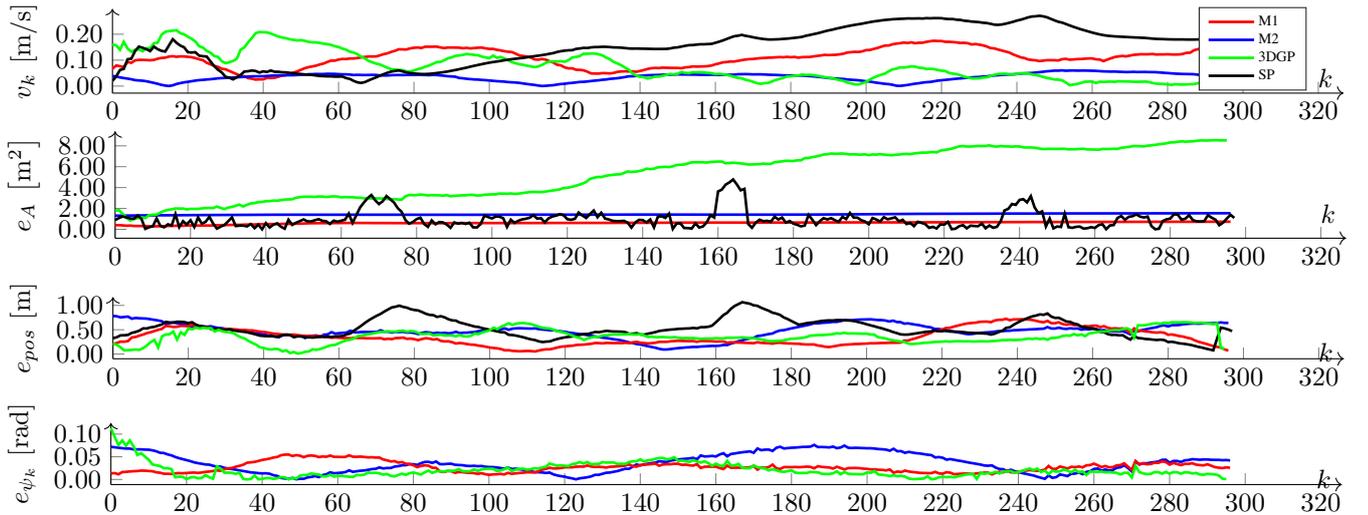
\begin{table}
	\centering
	\caption{The root mean squared errors for velocity $ \mathcal{E}_v $, area $ \mathcal{E}_A $, Cartesian position $ \mathcal{E}_{\mathtt{pos}} $ and orientation $ \mathcal{E}_\psi $ of the static scenario.} \label{tab:mean_static}
	\begin{tabular}{c|c|c|c|c}
		Method & $ \mathcal{E}_v $ & $ \mathcal{E}_A $ & $ \mathcal{E}_{\mathtt{pos}} $ & $ \mathcal{E}_\psi $ \\
		\hline
		M1 & 0.100 & \textbf{0.617} & 0.308 & 0.029 \\
		M2 & \textbf{0.027}&1.433&0.393&0.041\\
		3DGP&0.095&5.789&\textbf{0.304}&\textbf{0.026}\\
		SP&0.168&1.253&0.436& -
	\end{tabular}
\end{table}
\begin{table}
	\centering
	\caption{Comparison of the mean run time with 20 points $ \overline{t}_{20} $ and 50 points $ \overline{t}_{50} $ and bounding box $ \overline{t}_{\mathtt{SP}} $ as measurements.}\label{tab:time}
	\begin{tabular}{c|c|c|c}
		Method & $ \overline{t}_{20} [\SI{}{\milli\second}] $ & $ \overline{t}_{50}[\SI{}{\milli\second}]$ & $\overline{t}_{\mathtt{SP}}[\SI{}{\milli\second}]$ \\
		\hline
		M1 & 6.5 & 18.8&-  \\
		M2 & \textbf{3.0}&\textbf{3.1}&-\\
		3DGP&32.5&118.7&-\\
		\hline
		SP&-&-&0.8
	\end{tabular}
\end{table}
\Cref{fig:static} shows the estimates of the static vehicle at every time step $ k $.
It can be seen that our methods generate better or comparable results in every tested quantity.
Moreover, \Cref{tab:mean_static} supports this conclusion.
Furthermore, \Cref{tab:time} shows that, except for SP, M2 is the fastest version with only \SI{3.1}{\milli\second} per time step, followed by M1 with \SI{18.8}{\milli\second}.
3DGP needs \SI{118.7}{\milli\second} with 50 measurements.
Hence, it is not real-time capable with that many measurements.
\subsection{Scenario 2: Dynamic Object}
In this scenario, the shape of a dynamic vehicle has to be estimated alongside its dynamics, \eg Cartesian position and velocity.
Furthermore, 20 laser points are randomly sampled from the segmented object's point cloud and its 2D convex hull, as can be seen in \Cref{fig:car_points_back,fig:car_points_back_20}.
\subsubsection{Initialization}
Here, for all methods the initial shapes and most of the parameters are the same as in scenario 1.
Only the input noise is set to $ \bm{C}_{\dot{v}} = 0.2 $ and $ \bm{C}_{\dot{c}} = 0.05 $
, whereas $ Q_{w^{ij}} = 0.01 \;\forall i,j$.
\subsubsection{Evaluation}
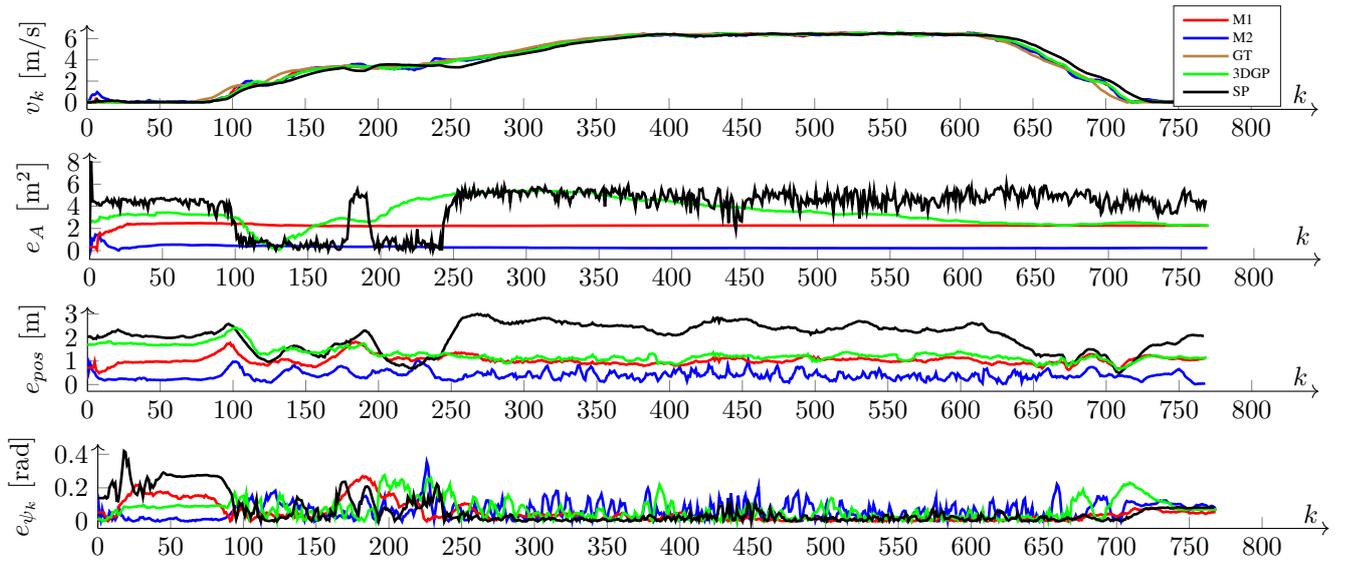
\begin{figure*}[t!]
	\centering
	\begin{tikzpicture}
\begin{axis}[
height=2.7cm,
width=\linewidth,
legend style={legend cell align=left, align=right, draw=white!15!black,nodes={scale=0.55, transform shape}},
legend pos = south east,
xlabel style={at={(0.999,0.4)},anchor=north east},
axis x line*=bottom,
axis y line*=left,
axis line style={->},
xmin=0,
xlabel=$ k $,
y unit=m/s,
ylabel=$ v_k$,
]

\addplot[color=red, line width=1pt] table[y expr={abs(\thisrow{gv})},x expr=\coordindex,col sep=comma]{nurbs_result_dynamic.csv};
\addplot[color=blue, line width=1pt] table[y expr={abs(\thisrow{gv})},x expr=\coordindex,col sep=comma]{surface_result_dynamic.csv};
\addplot[color=brown, line width=1pt] table[y expr={abs(\thisrow{ginsv})},x expr=\coordindex,col sep=comma]{surface_result_dynamic.csv};
\addplot[color=green, line width=1pt] table[y expr={abs(\thisrow{gv})},x expr=\coordindex,col sep=comma]{gp_result_dynamic.csv};
\addplot[color=black!60!black, line width=1pt] table[y expr={abs(\thisrow{gv})},x expr=\coordindex,col sep=comma]{point_result_dynamic.csv};
\legend{M1,M2,GT,3DGP,SP}
\end{axis}
\end{tikzpicture}
\\
\begin{tikzpicture}
\begin{axis}[
height=3cm,
width=\linewidth,
legend style={legend cell align=left, align=right, draw=white!15!black,nodes={scale=0.55, transform shape}},
legend pos = south east,
xlabel style={at={(0.999,0.4)},anchor=north east},
axis x line*=bottom,
axis y line*=left,
axis line style={->},
xmin=0,
xlabel=$ k $,
y unit=m^2,
ylabel=$ e_A$,
]
\addplot[color=blue, line width=1pt] table[y expr={abs((\thisrow{l}*\thisrow{w})-(\thisrow{insl}*\thisrow{insw})},x expr=\coordindex,col sep=comma]{surface_result_dynamic.csv};
\addplot[color=red, line width=1pt] table[y expr={abs((\thisrow{l}*\thisrow{w})-(\thisrow{insl}*\thisrow{insw})},x expr=\coordindex,col sep=comma]{nurbs_result_dynamic.csv};
\addplot[color=green, line width=1pt] table[y expr={abs((\thisrow{l}*\thisrow{w})-(\thisrow{insl}*\thisrow{insw})},x expr=\coordindex,col sep=comma]{gp_result_dynamic.csv};
\addplot[color=black!60!black, line width=1pt] table[y expr={abs((\thisrow{l}*\thisrow{w})-(\thisrow{insl}*\thisrow{insw})},x expr=\coordindex,col sep=comma]{point_result_dynamic.csv};
\end{axis}
\end{tikzpicture}
\\
\begin{tikzpicture}
\begin{axis}[
height=2.7cm,
width=\linewidth,
legend style={legend cell align=left, align=left, draw=white!15!black},
legend pos = south west,
xlabel style={at={(0.999,0.4)},anchor=north east},
axis x line*=bottom,
axis y line*=left,
axis line style={->},
xmin=0,
xlabel=$ k $,
y unit=m,
ylabel=$ e_{pos}$,
]
\addplot[color=blue, line width=1pt] table[y expr={abs(\thisrow{ginssx}-\thisrow{gx})+abs(\thisrow{ginssy}-\thisrow{gy}))},x expr=\coordindex,col sep=comma]{surface_result_dynamic.csv};
\addplot[color=red, line width=1pt] table[y expr={abs(\thisrow{ginssx}-\thisrow{gx})+abs(\thisrow{ginssy}-\thisrow{gy}))},x expr=\coordindex,col sep=comma]{nurbs_result_dynamic.csv};
\addplot[color=black!60!black, line width=1pt] table[y expr={abs(\thisrow{ginssx}-\thisrow{gx})+abs(\thisrow{ginssy}-\thisrow{gy}))},x expr=\coordindex,col sep=comma]{point_result_dynamic.csv};

\addplot[color=green, line width=1pt] table[y expr={abs(\thisrow{ginssx}-\thisrow{gx})+abs(\thisrow{ginssy}-\thisrow{gy}))},x expr=\coordindex,col sep=comma]{gp_result_dynamic.csv};
\end{axis}
\end{tikzpicture}
\\
\begin{tikzpicture}
\begin{axis}[
height=2.7cm,
width=\linewidth,
legend style={legend cell align=left, align=left, draw=white!15!black},
legend pos = south west,
xlabel style={at={(0.999,0.4)},anchor=north east},
axis x line*=bottom,
axis y line*=left,
axis line style={->},
xmin=0,
xlabel=$ k $,
y unit=rad,
ylabel=$ e_{\psi_k}  $,
]
\addplot[color=blue, line width=1pt] table[y expr={abs(\thisrow{gyaw}-\thisrow{ginsyaw})},x expr=\coordindex,col sep=comma]{surface_result_dynamic.csv};
\addplot[color=red, line width=1pt] table[y expr={abs(\thisrow{gyaw}-\thisrow{ginsyaw})},x expr=\coordindex,col
sep=comma]{nurbs_result_dynamic.csv};
\addplot[color=green, line width=1pt] table[y expr={abs(\thisrow{gyaw}-\thisrow{ginsyaw})},x expr=\coordindex,col sep=comma]{gp_result_dynamic.csv};
\addplot[color=black, line width=1pt] table[y expr={abs(\thisrow{gyaw}-\thisrow{ginsyaw})},x expr=\coordindex,col sep=comma]{point_result_dynamic.csv};

\end{axis}
\end{tikzpicture}
	\caption{Results of estimating a dynamic vehicle, with velocity $ v_k $, area error $ e_A $, Cartesian position error $ e_{pos} $ and orientation error $ e_\psi $. The area of the estimated shapes is defined as the product of the length and width of the encasing rectangle. Furthermore, one time step $ k $ is \SI{100}{\milli\second}.} \label{fig:dynamic}
\end{figure*}
\begin{table}
	\centering
	\caption{The root mean squared errors for velocity $ \mathcal{E}_v $, area $ \mathcal{E}_A $, Cartesian position $ \mathcal{E}_{\mathtt{pos}} $ and orientation $ \mathcal{E}_\psi $ of the dynamic scenario.}\label{tab:mean_dynamic}
	\begin{tabular}{c|c|c|c|c}
		Method & $ \mathcal{E}_v $ & $ \mathcal{E}_A $ & $ \mathcal{E}_{\mathtt{pos}} $ & $ \mathcal{E}_\psi $ \\
		\hline
		M1 & \textbf{0.196} & 2.259 & \textbf{0.303} & \textbf{0.076} \\
		M2 & 0.241&\textbf{0.323}&0.304&0.106\\
		3DGP&0.227&3.576&1.018&0.090\\ 
		SP&0.436&4.329&1.604& 0.101
	\end{tabular}
\end{table}
\Cref{fig:dynamic} shows the estimation process of the dynamic scenario.
As can be seen in \Cref{tab:mean_dynamic} our methods generate promising results in the dynamic scenario, too.
Only 3DGP has similar velocity and orientation results as our methods.
Moreover, M2 again is the fastest of the ETT approaches, which is shown in \Cref{tab:time}.
Furthermore, M1 only needs \SI{6.5}{\milli\second} with 20 measurements per time step and 3DGP also is real-time capable with a mean run time of \SI{32.5}{\milli\second}.
The estimated surfaces after a couple of time steps, where the measurements only originated from the back of the car are shown in \Cref{fig:quali_dynamic}.
\subsection{Limitations of the Approach}
In the evaluation and development of this approach, we discovered that a fix process noise for the shape states has difficulties with a large error in the initialization of the velocity state.
On the one hand, with a low process noise the velocity will be correctly estimated but the shape estimation needs more cycles to converge. On the other hand, with a high process noise the shape estimation has a fast convergence, but the velocity state needs more cycles to converge.
Another limitation of M2 is that only approximately cuboid shaped objects like vehicles, persons and bicycles can be estimated.
\section{conclusion} \label{sec:conclusion}
This paper proposed a novel 3D shape model, based on NURBS surfaces, for simultaneously estimating an extended target's unknown 3D shape and dynamics.
It has been shown, that the estimation capability is above or comparable to state-of-the-art ETT approaches.
Evaluations have been done in real-world scenarios with a high-resolution 3D LiDAR, which is a popular sensor of autonomous cars at research institutes.
Furthermore, the proposed methods have shown real-time capability with the fastest method's mean run-time of \SI{3.0}{\milli\second}.
Future work will focus on applying it to multi-target tracking scenarios as well as adaptive process noise for the shape part to account for wrong state initialization.
%

\section*{ACKNOWLEDGMENT}
The authors gratefully acknowledge funding by the Federal Office of Bundeswehr Equipment, Information Technology and In-Service Support (BAAINBw).

\bibliographystyle{IEEEtran}
\bibliography{IEEEabrv,additional_abrv,et_al,tas_papers,bena}

\begin{thebibliography}{10}
\providecommand{\url}[1]{#1}
\csname url@samestyle\endcsname
\providecommand{\newblock}{\relax}
\providecommand{\bibinfo}[2]{#2}
\providecommand{\BIBentrySTDinterwordspacing}{\spaceskip=0pt\relax}
\providecommand{\BIBentryALTinterwordstretchfactor}{4}
\providecommand{\BIBentryALTinterwordspacing}{\spaceskip=\fontdimen2\font plus
\BIBentryALTinterwordstretchfactor\fontdimen3\font minus
  \fontdimen4\font\relax}
\providecommand{\BIBforeignlanguage}[2]{{%
\expandafter\ifx\csname l@#1\endcsname\relax
\typeout{** WARNING: IEEEtran.bst: No hyphenation pattern has been}%
\typeout{** loaded for the language `#1'. Using the pattern for}%
\typeout{** the default language instead.}%
\else
\language=\csname l@#1\endcsname
\fi
#2}}
\providecommand{\BIBdecl}{\relax}
\BIBdecl

\bibitem{bena:overviewGT}
L.~Mihaylova, A.~Y. Carmi \emph{et~al.}, ``{Overview of Bayesian Sequential
  Monte Carlo Methods for Group and Extended Object Tracking},'' \emph{Digital
  Signal Processing}, 2014.

\bibitem{bena:Granstroen2016}
K.~Granströn, M.~Baum, and S.~Reuter, ``{Extended Object Tracking:
  Introduction, Overview and Applications},'' \emph{arXiv preprint
  arXiv:1604.00970}, 2016.

\bibitem{bena:Piegl1991}
L.~Piegl, ``{On NURBS: A Survey},'' \emph{{IEEE} Comput. Graph. Appl.}, 1991.

\bibitem{bena:Baum2012}
M.~Baum, F.~Faion, and U.~D. Hanebeck, ``{Modeling the Target Extent with
  Multiplicative Noise},'' in \emph{{Proc. Int. Conf. Information Fusion
  (FUSION)}}, 2012.

\bibitem{bena:Gilholm2005}
K.~Gilholm and D.~Salmond, ``{Spatial Distribution Model for Tracking Extended
  Objects},'' \emph{IEEE Proceedings-Radar, Sonar and Navigation}, 2005.

\bibitem{bena:Granstroem2011}
K.~Granström, C.~Lundquist, and U.~Orguner, ``{Tracking Rectangular and
  Elliptical Extended Targets using Laser Measurements},'' in \emph{{Proc. Int.
  Conf. Information Fusion (FUSION)}}, 2011.

\bibitem{bena:lmbextended}
M.~Beard, S.~Reuter \emph{et~al.}, ``{Multiple Extended Target Tracking with
  Labeled Random Finite Sets},'' \emph{{IEEE} Trans. Signal Process.}, vol.~64,
  no.~7, 2016.

\bibitem{bena:Feldmann2011}
M.~Feldmann, D.~Franken, and W.~Koch, ``{Tracking of Extended Objects and Group
  Targets using Random Matrices},'' \emph{{IEEE} Trans. Signal Process.}, 2011.

\bibitem{bena:Baum2014}
M.~Baum and U.~D. Hanebeck, ``{Extended Object Tracking with Random
  Hypersurface Models},'' \emph{IEEE Trans. on Aerospace and Electronic
  Systems}, 2014.

\bibitem{LevelSetRHM}
A.~Zea, F.~Faion, M.~Baum, and U.~D. Hanebeck, ``{Level-Set Random Hypersurface
  Models for Tracking Non-Convex Extended Objects},'' in \emph{{Proc. Int.
  Conf. Information Fusion (FUSION)}}, 2013.

\bibitem{bena:wahlstromextended}
N.~Wahlstr{\"o}m and E.~{\"O}zkan, ``{Extended Target Tracking using Gaussian
  Processes},'' \emph{{IEEE} Trans. Signal Process.}, vol.~63, no.~16, 2015.

\bibitem{bena:Martin2017}
M.~Michaelis, P.~Berthold, D.~Meissner, and H.-J. Wuensche, ``{Heterogeneous
  Multi-Sensor Fusion for Extended Objects in Automotive Scenarios using
  Gaussian processes and a GMPHD-filter},'' in \emph{Sensor Data Fusion:
  Trends, Solutions, Applications (SDF)}.\hskip 1em plus 0.5em minus
  0.4em\relax IEEE, 2017.

\bibitem{bena:Kumru2018}
M.~Kumru and E.~Özkan, ``{3D Extended Object Tracking Using Recursive Gaussian
  Processes},'' in \emph{{Proc. Int. Conf. Information Fusion (FUSION)}}, 2018.

\bibitem{bena:Faion2015}
F.~Faion, A.~Zea \emph{et~al.}, ``{Recursive Bayesian Pose and Shape Estimation
  of 3D Objects using Transformed Plane Curves},'' in \emph{Sensor Data Fusion:
  Trends, Solutions, Applications (SDF)}.\hskip 1em plus 0.5em minus
  0.4em\relax IEEE, 2015.

\bibitem{bena:Steinemann2012}
P.~Steinemann, J.~Klappstein \emph{et~al.}, ``{3D Outline Contours of Vehicles
  in 3D-Lidar-Measurements for Tracking Extended Targets},'' in \emph{{Proc.
  IEEE Intelligent Vehicles Symp. (IV)}}, 2012.

\bibitem{bena:Schreier2016}
M.~Schreier, V.~Willertz, and J.~Adamy, ``{Compact Representation of Dynamic
  Driving Environments for ADAS by Parametric Free Space and Dynamic Object
  Maps},'' \emph{{IEEE} Trans. Intell. Transp. Syst.}, 2016.

\bibitem{bena:burger_lane_2019_itsc}
P.~Burger, B.~Naujoks, and H.-J. Wuensche, ``{Unstructured Road SLAM using Map
  Predictive Road Tracking},'' in \emph{{Proc. IEEE Intelligent Transportation
  Syst. Conf. (ITSC)}}, 2019.

\bibitem{bena:Zea2016}
A.~Zea, F.~Faion, and U.~D. Hanebeck, ``{Tracking Elongated Extended Objects
  using Splines},'' in \emph{{Proc. Int. Conf. Information Fusion (FUSION)}},
  2016.

\bibitem{bena:Daniyan2018}
A.~Daniyan, S.~Lambotharan \emph{et~al.}, ``{Bayesian Multiple Extended Target
  Tracking Using Labeled Random Finite Sets and Splines},'' \emph{{IEEE} Trans.
  Signal Process.}, 2018.

\bibitem{bena:Kaulbersch2018}
H.~Kaulbersch, J.~Honer, and M.~Baum, ``{A Cartesian B-Spline Vehicle Model for
  Extended Object Tracking},'' in \emph{{Proc. Int. Conf. Information Fusion
  (FUSION)}}, 2018.

\bibitem{bena:Schuster2008}
R.~Schuster, E.~Richter, and G.~Wanielik, ``{Comparison and Evaluation of
  Advanced Motion Models for Vehicle Tracking},'' in \emph{{Proc. Int. Conf.
  Information Fusion (FUSION)}}, 2008.

\bibitem{bena:Faion2012}
F.~Faion, M.~Baum, and U.~D. Hanebeck, ``{Tracking 3D Shapes in Noisy Point
  Clouds with Random Hypersurface Models},'' in \emph{{Proc. Int. Conf.
  Information Fusion (FUSION)}}, 2012.

\bibitem{UKFMerwe}
E.~A. Wan and R.~van~der Merwe, ``{The Unscented Kalman Filter for Nonlinear
  Estimation},'' in \emph{Proceedings of the IEEE Adaptive Systems for Signal
  Processing, Communications, and Control Symposium}, Lake Louise, AB, Canada,
  2000.

\bibitem{bena:Saini2015}
D.~Saini, S.~Kumar, and T.~R. Gulati, ``{Reconstruction of free-form space
  curves using NURBS-snakes and a quadratic programming approach},''
  \emph{Computer Aided Geometric Design}, 2015.

\bibitem{bena:Goldman2005}
R.~Goldman, ``{Curvature formulas for implicit curves and surfaces},''
  \emph{Computer Aided Geometric Design}, 2005.

\bibitem{bena:GDPF}
B.~Naujoks, P.~Burger, and H.-J. Wuensche, ``{The Greedy Dirichlet Process
  Filter - An Online Clustering Multi-Target Tracker},'' in \emph{IEEE Global
  Conference on Signal and Information Processing (GlobalSIP)}, 2018.

\bibitem{bena:burger_iv2018}
P.~Burger and H.-J. Wuensche, ``{Fast Multi-pass 3D Point Segmentation Based on
  a Structured Mesh Graph for Ground Vehicles},'' in \emph{{Proc. IEEE
  Intelligent Vehicles Symp. (IV)}}, Jun. 2018.

\bibitem{bena:burger_itsc2018}
P.~Burger, B.~Naujoks, and H.-J. Wuensche, ``{Fast Dual Decomposition based
  Mesh-Graph Clustering for Point Clouds},'' in \emph{{Proc. IEEE Intelligent
  Transportation Syst. Conf. (ITSC)}}, 2018.

\bibitem{bena:BeNaIV2018}
B.~Naujoks and H.-J. Wuensche, ``{An Orientation Corrected Bounding Box Fit
  Based on the Convex Hull under Real Time Constraints},'' in \emph{{Proc. IEEE
  Intelligent Vehicles Symp. (IV)}}, 2018.

\end{thebibliography}
\end{document}